\documentclass[twocolumn,showpacs,preprintnumbers,amsmath,amssymb,prb]{revtex4}
\usepackage{amssymb}
\usepackage{amsmath}
\usepackage{txfonts}
\usepackage{stmaryrd}
\usepackage{graphicx}
\usepackage{dcolumn}
\usepackage{bm}
\usepackage{subfigure}

\begin{document}

\title{A new method to find full complex roots of a complex dispersion equation for light propagation}

\author{Li Wan}
\email{liwan_china@yahoo.com.cn} \affiliation{Department of Physics,
Wenzhou University, Wenzhou 325035, People$'$s Republic of China}


\begin{abstract}
A new numerical method is presented to find full complex roots of a
complex dispersion equation. For the application of the solution,
the complex dispersion equation of a cylindrical metallic nanowire
is investigated. By using this method, locus of Brewster angle,
complex dispersion curves of Surface Plasmon Polaritons (SPPs) and
complex bulk modes can be obtained in once calculation. Approximate
analytical solution to the complex dispersion equation has also been
derived to verify our method.
\end{abstract}

\maketitle
\section{Introduction}

Dispersion relation is the basic property of light propagating in
mediums, which specifies the relation between the wavevector $k$ and
the frequency  $\omega$ of the light. Recently, the light
propagation in metallic nanostructures has attracted enormous
attentions from researchers due to the excitation of the Surface
Plasmon-polaritons(SPPs)
~\cite{Barnes1,Maier1,Ozbay1,Lal1,Gramotnev1} by the light. The SPPs
are known for the wide application in nano optics attributed to the
spatial localization of the SPPs at the metal-dielectric interfaces,
which can be guided to manipulate light in nanoscaled photonic
circuitry~\cite{Barnes1,Maier1,Ozbay1,Lal1,Gramotnev1}. Furthermore,
the spatial confinement of the SPPs can enhance the field intensity
at the interfaces, which can influence the luminescence
intensities~\cite{Talley1,Prodan1,Kelly1} and life time of emitters
close to the interfaces.~\cite{Abajo1,Chicanne1} In order to
understand the physical properties of SPPs in plasmonic structures,
it is important to get the dispersion relations of light propagation
in the plasmonic structures. Principally, the dispersion relations
can be obtained by solving the Maxwell equations with the boundary
conditions of the plasmonic structures imposed. For simple plasmonic
structures, such as planar metal surfaces, the dispersion of SPPs
can be solved
analytically.~\cite{Ruppin1,Archambault1,Halevi1,Rice1} But for the
most structures which are nonsymmetrical and irregular, codes such
as finite-element method (FEM) and finite-difference time-domain
techniques (FDTD) need to be applied for numerical calculations. The
basic technique of FEM or FDTD is to discrete the Maxwell equations,
which are known for the time consumption. Fortunately, for some
symmetric structures, such as cylindrical
nanowires,~\cite{Ruppin1,Pfeiffer1,Ashley1,Chang1,Chen1,Novotnv1}
the Maxwell equations imposed with the boundary conditions can be
transformed into a dispersion equation. Solving the dispersion
equation can improve the efficiency to get the dispersion relation
rather than the numerical calculation by the discretion of Maxwell
equations.

The dispersion equation normally is nonlinear and transcendental.
What is more, when the metal loss is introduced into the metal
dielectric,~\cite{Halevi1,Rice1,Arakawa1,Alexander1} the dispersion
equation then is a complex and transcendental equation. Thus, the
dispersion relations obtained from the complex equation are also
complex. There exist two types of the complex dispersion relations
for one same plasmonic structure, specifying either a complex
frequency as a function of a real wave vector, noted as
$complex-\omega$ for convenience, or a complex wave vector as a
function of a real frequency, noted as $complex-k$. The two types of
dispersion relations are rather different, even though they are for
one same structure. The back bending as a characteristic of the
$complex-k$ relation is absent in the $complex-\omega$ relation. The
latter is an asymptotic curve.~\cite{Chen1,Novotnv1} It has been
suggested that the $complex-k$ solution of the dispersion relation
describes the SPPs mode decaying spatially while the
$complex-\omega$ solution is for the SPPs decaying in time rather
than in space.~\cite{Archambault1,Udagedara1,Yao1} The discrepancy
of the two solutions has been considered to be originated from the
metal Ohmic loss.~\cite{Halevi1,Rice1,Arakawa1,Alexander1} Such
conclusion is originally drawn from the study of SPPs on planar
metallic-dielectric surfaces. When a perfect metal without damping
is considered in this planar case, the two types of solutions of the
SPPs dispersion relations overlap with a same asymptotic behavior.
When the metal Ohmic loss is introduced into the dielectric response
of metal, the two types of solutions are then different. The
original of the discrepancy of the two solutions due to the metal
loss has also been confirmed in cylindrical metallic
nanowires.~\cite{Wan1}

The main mission to get the dispersion relation of the light
propagation in plasmonic structures then is to find the complex
roots of the equation. The complex dispersion equation can be
described by $f(x,y,z)=0$ with three real variables $x$, $y$ and
$z$. For example, in the $complex-\omega$ solution, $x$, $y$ and $z$
represent the real part of $\omega$ ($Re[\omega]$), the imaginary
part of $\omega$ ($Im[\omega]$) and real wavevector $k$
respectively. In the $complex-k$ solution, $x$, $y$ and $z$ then
represent the real part of $k$ ($Re[k]$), imaginary part of $k$
($Im[k]$) and real $\omega$ respectively. To solve this complex
equation, normally one variable is given, say $z$, then the complex
equation can be simplified to be a complex equation $f(x,y)=0$ with
only two real variables. One commonly used method to solve this
equation is to choose all possible values of $x$ and $y$ in their
given ranges to check if they satisfy the equation or not. To
achieve this purpose, a coordination system with $x$ and $y$ axis is
gridded with an enough small mesh size and then the values of $x$
and $y$ at each grid point are substituted into the equation.
Considering the error allowance $\delta$, the criterion for this
grid method is to find the roots $(x_{0},y_{0})$ if they satisfy
$|f(x_{0},y_{0})|<\delta$. To this grid method, there exit two main
short comings. The first is that in order to get the full solutions
to the equation, one needs to gird the coordinate system with an
enough small mesh size. Or, the solutions may be lost. However, in
the reality for the calculation, we find that the mesh size can
reach the order of the magnitude of $10^{-15}$ and even smaller,
which increases the computation time. The second shortcoming is that
the criterion can not guarantee the  $(x_{0},y_{0})$ to be the root
even though they satisfy the criterion since $\delta$ is not
rigorously equal to zero. No matter how small the $\delta$ is taken,
the second shortcoming still remains since $\delta\neq 0$. To
improve the computation efficiency, the alternative method is the
Newton-Raphson (NR) method, which can converge to the roots of the
equation quickly. However, it is well known that the NR method may
miss the roots if there exist multiroots of the equation. What is
more, the NR method may not converge if the initial estimate is not
close enough to the root, or may converge to wrong root. Especially
for the complex transcendental equation, which is very common for
the light propagation in plasmonic structures when the metal Ohmic
loss is introduced, the NR method will lose its power to find the
roots. In this paper, we propose one new method to find full complex
roots of complex transcendental equations. As an example for the
application, the dispersion equation of light propagating in a
cylindrical metallic nanowire is investigated.

\section{method}
\subsection{linear approximation}
\begin{figure}[t!]
\begin{center}
\includegraphics[width=2.0in, height=2.7in, angle=270]{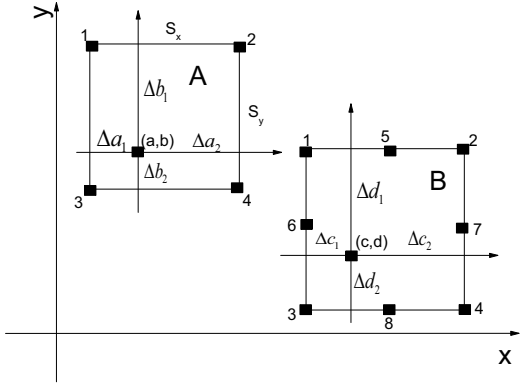}
\vspace*{-0.2cm}
\end{center}
\caption{Schematic of our method to get full complex roots of a
complex dispersion equation. $S_x$ and $S_y$ are the mesh sizes
along $x$ and $y$ axis respectively. $(a,b)$ and $(c,d)$ are assumed
to be the roots of the equation, enclosed in mesh A and mesh B
respectively. The former is used for the linear approximation, while
the latter is for the parabola  approximation. } \label{fig.1}
\end{figure}
The main mission of our method to solve the complex equation
$f(x,y)=0$ is to find criterions used to judge if the values of the
variables are the roots or not. Similarly to the grid method, the
coordination system with $x$ and $y$ axis is gridded firstly.
Suppose that $a$ and $b$ are the roots of the equation satisfying
$f(a,b)=0$ and the point $(a,b)$ is enclosed in the mesh
A(fig.~\ref{fig.1}), then the function at each point of the mesh
corner can be expanded at the point $(a,b)$  by the first order
Taylor series as
\begin{subequations}
\label{eq.1}
\begin{equation}
\label{eq.1a} f(x,y)_1=f(a,b)+f_{x}'(a,b)(-\triangle
a_1)+f_{y}'(a,b)(\triangle b_1),
\end{equation}
\begin{equation}
\label{eq.1b} f(x,y)_2=f(a,b)+f_{x}'(a,b)(\triangle
a_2)+f_{y}'(a,b)(\triangle b_1),
\end{equation}
\begin{equation}
\label{eq.1c} f(x,y)_3=f(a,b)+f_{x}'(a,b)(-\triangle
a_1)+f_{y}'(a,b)(-\triangle b_2),
\end{equation}
\begin{equation}
\label{eq.1d} f(x,y)_4=f(a,b)+f_{x}'(a,b)(\triangle
a_2)+f_{y}'(a,b)(-\triangle b_2).
\end{equation}
\end{subequations}
Here, subscript $j$ of $f(x,y)_{j(j=1,2,3,4)}$ represents the
calculated results of the function $f(x,y)$ at the corner point
$j(j=1,2,3,4)$ of mesh A, which has been labeled in the
fig.~\ref{fig.1}. The complex $f_{x(y)}'(a,b)$ is the partial
derivative of the function with respect to $x(y)$ at the point of
$(a,b)$. $\triangle a_1$, $\triangle a_2$, $\triangle b_1$, and
$\triangle b_2$ are the absolute values of the coordination
components of the four corner points of the mesh A with respect to
the point $(a,b)$ considered as the coordinate origin.
Eq.(\ref{eq.1a}) subtracted from eq.(\ref{eq.1b}) gives
\begin{subequations}
\label{eq.2}
\begin{equation}
\label{eq.2a} f_{x}'(a,b)=\frac{f(x,y)_2-f(x,y)_1}{\triangle
a_1+\triangle a_2}=\frac{f(x,y)_2-f(x,y)_1}{S_x},
\end{equation}
and eq.(\ref{eq.1a}) subtracted from eq.(\ref{eq.1c}) gives
\begin{equation}
\label{eq.2b} f_{y}'(a,b)=\frac{f(x,y)_1-f(x,y)_3}{\triangle
b_1+\triangle b_2}=\frac{f(x,y)_1-f(x,y)_3}{S_y}.
\end{equation}
\end{subequations}
The $S_x$ and $S_y$ are the mesh sizes along the $x$ and $y$ axis
respectively, which are given as constants when gridding the
coordination system. We substitute eq.(\ref{eq.2}) into
eq.(\ref{eq.1a}), obtaining
\begin{subequations}
\label{eq.3}
\begin{equation}
\label{eq.3a} \triangle
a_1=\frac{L_{2}L_{12}-L_{1}L_{22}}{L_{21}L_{12}-L_{11}L_{22}},
\end{equation}
\begin{equation}
\label{eq.3b} \triangle
b_1=\frac{L_{1}L_{21}-L_{2}L_{11}}{L_{21}L_{12}-L_{11}L_{22}},
\end{equation}
\end{subequations}
with
\begin{eqnarray}
\begin{array}{lll}
L_{1}=Re[f(x,y)_1], & L_{2}=Im[f(x,y)_1], & L_{11}=Re[-f_{x}'(a,b)],\nonumber\\
L_{12}=Re[f_y'(a,b)], & L_{21}=Im[-f_x'(a,b)], &
L_{22}=Im[f_y'(a,b)].\nonumber
\end{array}
\end{eqnarray}
Similarly, by substituting eq.(\ref{eq.2}) into eq.(\ref{eq.1b}),
(\ref{eq.1c}) and (\ref{eq.1d}), we can get two sets of the
$\triangle a_1$, $\triangle a_2$, $\triangle b_1$, and $\triangle
b_2$, noted as $\triangle a_{m(m=1,2)}^{n(n=1,2)}$, and $\triangle
b_{m(m=1,2)}^{n(n=1,2)}$ for convenience. Now we give the criterion
to find the roots of the equation. If the $a$ and $b$ are the roots
of the complex equation satisfying $f(a,b)=0$, then
\begin{subequations}
\label{eq.4}
\begin{eqnarray}
\label{eq.4a} \left \{
\begin{array}{ll}
 \triangle a_{m(m=1,2)}^{n(n=1,2)}>0,  & \triangle b_{m}^{n}>0,\\
 \\
 \triangle a_{m(m=1,2)}^{1}=\triangle a_{m(m=1,2)}^{2},  & \triangle b_{m(m=1,2)}^{1}=\triangle
 b_{m(m=1,2)}^{2},
 \end{array}
  \right.
\end{eqnarray}
must be held. Or, there exists no root in the grid A due to the
$f(a,b)\neq 0$ in eq.(\ref{eq.1}). Considering the error allowance
$\delta$, criterion (\ref{eq.4a}) can be modified as
\begin{eqnarray}
\label{eq.4b} \left \{
\begin{array}{ll}
 \triangle a_{m(m=1,2)}^{n(n=1,2)}>0,  & \triangle b_{m}^{n}>0,\\
 \\
 \left | \frac{\triangle a_{m(m=1,2)}^{1}-\triangle a_{m(m=1,2)}^{2}}{\triangle a_{m(m=1,2)}^{2}}\right
 |<\delta,
 & \left | \frac{\triangle b_{m(m=1,2)}^{1}-\triangle b_{m(m=1,2)}^{2}}{\triangle b_{m(m=1,2)}^{2}}\right |
 <\delta.
  \end{array}
  \right.
\end{eqnarray}
\end{subequations}
By using this method, the correct mesh enclosing the root can be
found. And then we can further gird the correct mesh and consider
the grid point with the minimum value of $|f(x,y)|$ in the mesh as
the root of the equation, at least they are very close to the roots.
We note that in our method the criterion $|f(x_0,y_0)|<\delta$ is
not fatal for the determination of the roots. And the function
values $|f(x,y)|$ at the points considered as the roots may be
larger than the error allowance $\delta$ if one wants to save
computation time. The criterion (\ref{eq.4}) can guarantee the
points close to the exact roots if the points are enclosed in the
correct meshes satisfying the criterion (\ref{eq.4}). In our method,
all the meshes are independent to each other. Thus, all possible
roots can be found, overcoming the problem met by the NR method.
What is more, this method focuses on the finding of correct meshes
enclosing the roots instead of the finding of correct points
matching the roots, which saves much more computation time.

\subsection{parabola approximation}
Generally, if the derivative $f_{x(y)}'(x,y)$ in eq.(\ref{eq.1}) may
equal to zero when the function $f(x,y)$ has a parabola-like shape,
we can expand the function with the Taylor series up to order two
\begin{eqnarray}
\label{eq.5} f(x,y)=&& f(c,d)+f_{1x}(c,d)\triangle
c+f_{1y}(c,d)\triangle
d\nonumber \\
&& +f_{2xx}(c,d)\triangle c^{2}+f_{2yy}(c,d)\triangle
d^{2}+f_{2xy}(c,d)\triangle c\triangle d.\nonumber\\
&&
\end{eqnarray}
Here, $f_{1x(y)}$ is the coefficient for the first order Taylor
expansion and the $f_{2x(y)x(y)}$ is the coefficient for the second
order. If the $c$ and $d$ are the roots of the complex equation,
then $f(c,d)=0$ is held. For illustration, point $(c,d)$ is enclosed
in the mesh B, shown in the fig.~\ref{fig.1}. Each point of mesh B
used for the calculation has been labeled by $j(j=1,2,3,4,5,6,7,8)$
in the fig.1. Then, we can get
\begin{subequations}
\label{eq.6}
\begin{equation}
\label{eq.6a}
f_{2xy}(c,d)=\frac{f(x,y)_2+f(x,y)_3-f(x,y)_1-f(x,y)_4}{S_xS_y},
\end{equation}
and two sets of the $f_{2xx}(c,d)$ and $f_{2yy}(c,d)$, noted as
$f_{2xx}^{n(n=1,2)}(c,d)$ and
$f_{2yy}^{n(n=1,2)}(c,d)$:\\
\begin{eqnarray}
\label{eq.6b} \left \{
\begin{array}{l}
 f_{2xx}^{1}(c,d)=\frac{2(f(x,y)_1+f(x,y)_2)-4f(x,y)_5}{S_x^2},\\
 \\
 f_{2yy}^{1}(c,d)=\frac{2(f(x,y)_3+f(x,y)_1)-4f(x,y)_6}{S_y^2},
\end{array}
\right .
\end{eqnarray}
and
\begin{eqnarray}
\label{eq.6c} \left \{
\begin{array}{l}
 f_{2xx}^{2}(c,d)=\frac{2(f(x,y)_3+f(x,y)_4)-4f(x,y)_8}{S_x^2},\\
 \\
 f_{2yy}^{2}(c,d)=\frac{2(f(x,y)_2+f(x,y)_4)-4f(x,y)_7}{S_y^2}.
\end{array}
\right .
\end{eqnarray}
\end{subequations}
Here, the subscript $j$ of $f(x,y)_j$ specifies the function
$f(x,y)$ calculated at the point $j$ of the mesh B. After some
algebra, we get the following linear equation system:
\begin{subequations}
\label{eq.7}
\begin{eqnarray}
\label{eq.7a} f_{1x}(c,d)-2f_{2xx}(c,d)\triangle
c_1+f_{2xy}(c,d)\triangle
d_1 \nonumber \\
=\frac{4f(x,y)_5-3f(x,y)_1-f(x,y)_2}{S_x},
\end{eqnarray}
\begin{eqnarray}
\label{eq.7b} -f_{1y}(c,d)-2f_{2yy}(c,d)\triangle
d_1+f_{2xy}(c,d)\triangle
c_1 \nonumber \\
=\frac{4f(x,y)_6-3f(x,y)_1-f(x,y)_3}{S_y}.
\end{eqnarray}
\end{subequations}
The eq.(\ref{eq.7a}) and (\ref{eq.7b}) are complex. Each equation
can be separated into two equations with respect to the real and
imaginary parts of the equation. Now we consider three situations.
The first situation is $f_{1x}(c,d)=0$. The second situation is
$f_{1y}(c,d)=0$. And the third is both $f_{1x}(c,d)=0$ and
$f_{1y}(c,d)=0$. For the first situation, we can substitute
eq.(\ref{eq.6}) into the eq.(\ref{eq.7a}) and get two sets of
$\triangle c_1$ and $\triangle d_1$, noted as $\triangle
c_1^{p(p=1,2)}$ and $\triangle d_1^{p(p=1,2)}$. And for the second
situation, by substituting eq.(\ref{eq.6}) into eq.(\ref{eq.7b}), we
also obtain two sets of$\triangle c_1$ and $\triangle d_1$, noted as
$\triangle c_1^{p(p=3,4)}$ and $\triangle d_1^{p(p=3,4)}$. Last, for
the third situation, we then can get four sets of the roots
$\triangle c_1^{p(p=1,2,3,4)}$ and $\triangle d_1^{p(p=1,2,3,4)}$
from the eq.(\ref{eq.7}). Now we give the criterion to find roots of
the equation. If the $c$ and $d$ are the roots of the complex
equation satisfying $f(c,d)=0$ , then for the eq.(\ref{eq.6}) the
criterion
\begin{subequations}
\label{eq.8}
\begin{eqnarray}
\label{eq.8a} \left \{
\begin{array}{l}
\left |\frac{f_{2xx}^1-f_{2xx}^2}{f_{2xx}^1} \right | < \delta,\\
\\
\left |\frac{f_{2yy}^1-f_{2yy}^2}{f_{2yy}^1} \right | < \delta,
\end{array}
\right.
\end{eqnarray}
must be held. And for the first situation of $f_{1x}(c,d)=0$,
additional criterion is
\begin{eqnarray}
\label{eq.8b} \left \{
\begin{array}{l}
\left |\frac{\triangle c_1^1-\triangle c_1^2}{\triangle c_1^1} \right | < \delta,\\
\\
\left |\frac{\triangle d_1^1-\triangle d_1^2}{\triangle d_1^1}
\right | < \delta.
\end{array}
\right.
\end{eqnarray}
For the second situation of $f_{1y}(c,d)=0$, the additional
criterion is
\begin{eqnarray}
\label{eq.8c} \left \{
\begin{array}{l}
\left |\frac{\triangle c_1^3-\triangle c_1^4}{\triangle c_1^3} \right | < \delta,\\
\\
\left |\frac{\triangle d_1^3-\triangle d_1^4}{\triangle d_1^3}
\right | < \delta.
\end{array}
\right.
\end{eqnarray}
\end{subequations}
And for the third situation of both $f_{1x}(c,d)=0$ and
$f_{1y}(c,d)=0$, the additional criterions of eq.(\ref{eq.8b}) and
(\ref{eq.8c}) both should be held.

\section{calculation and discussion}
\subsection{dispersion equation}
In order to verify our method, the complex dispersion equation of
SPPs on the planar metallic surface has been investigated. The
rigorous analytical solutions to the dispersion
equation~\cite{Ruppin1,Archambault1,Halevi1,Rice1}
 in that case have confirmed the
validity of our method. However, in this paper we focus our
attention on the solutions to the dispersion equation of a
cylindrical metallic nanowire. The nanowire has the shape with the
radius of $r$ and an infinite length in a medium. The
electromagnetic field can be expanded with cylindrical harmonics. By
solving the Maxwell equation with the boundary conditions imposed,
the dispersion equation can be obtained as the following
transcendental equation:
\begin{eqnarray}
\label{eq.9}
\scriptsize{
\begin{vmatrix}
H_{n}^{(1)}(k_{r0}r)k_{r0}^{2} & 0 &-J_{n}(k_{r1}r)k_{r1}^{2} & 0\\
0 & H_{n}^{(1)}(k_{r0}r)k_{1}k_{r0}^{2} & 0 &-J_{n}(k_{r1}r)k_{0}k_{r1}^{2}\\
k_{r0}rH_{n}^{(1)\prime }(k_{r0}r)k_{0}k_{1} &
-nk_{z}H_{n}^{(1)}(k_{r0}r)k_{1}
& -k_{r1}rJ_{n}^{\prime }(k_{r1}r)k_{0}k_{1} & nk_zJ_{n}(k_{r1}r)k_{0}\\
-nk_{z}H_{n}^{(1)}(k_{r0}r) & k_{r0}rH_{n}^{(1)\prime}(k_{r0}r)k_{0}
& nk_{z}J_{n}(k_{r1}r) & -k_{r1}rJ_{n}^{\prime }(k_{r1}r)k_{1}
\end{vmatrix}=0},\nonumber \\
\end{eqnarray}
which repeats the reported results.~\cite{Pfeiffer1,Chang1} Here,
$H_{n}^{(1)}$ is the first kind of Hankel function with the order of
integer $n$ and $J_{n}$ is the Bessel function of $n$th order. The
Hankel and Bessel functions with denote represent the first order
differentiation. $k_{j(j=0,1)}$ is the wave vector with the value of
$k_{j}=\sqrt{\varepsilonup_j}\omega/c$, where $\varepsilonup_j$is
the dielectric function and the subscript $j$ labels the quantities
outside the nanowire ($j=0$)  or inside it ($j=1$). $c$ is the speed
of light in vacuum. The dielectric function of the metal can be
expressed as
\begin{equation}
\label{eq.10}
\varepsilonup_{1}(\omega)=\varepsilonup_{\infty}
[1-\frac{\omega_{p}^{2}}{\omega(\omega+i\tau)}],
\end{equation}
where $\omega_{p}$  is the bulk-plasmon frequency and $\tau$  is the
bulk electron relaxation rate,~\cite{Chen1} which reflects the metal
Ohmic loss. Eq.(\ref{eq.9}) then is a complex equation with $\tau$
introduced into the dielectric. $\varepsilonup_{\infty}$ in
eq.(\ref{eq.10}) is a constant for the general description of the
dielectric function of the metal. $k_z$ is the component of the SPPs
wave vector along the cylinder axial and the radial components of
the wave vectors are defined as
$k_{rj}=\!\sqrt{k_{j}^{2}-k_{z}^{2}}$. We note that the value of
$k_{r0}$ should be chosen to guarantee the imaginary part of
$k_{r0}$ to be positive since the light intensity should be decaying
away from the metal cylinder. The dispersion relation between $k_z$
and $\omega$ then can be obtained from the eq.(\ref{eq.9})
numerically. For the calculation, we renormalize the $\omega$ and
$\tau$ by $\omega_p$. Wave vector components of $k_{j(j=0,1)}$ ,
$k_z$ and $k_{rj(j=0,1)}$ are renormalized by $\omega_{p}/c$ and $r$
is renormalized by $c/\omega_{p}$.

\subsection{$complex-\omega$ solution}
\begin{figure*}[t!]
\centering
\includegraphics[width=2.0in, height=2.7in,
angle=270]{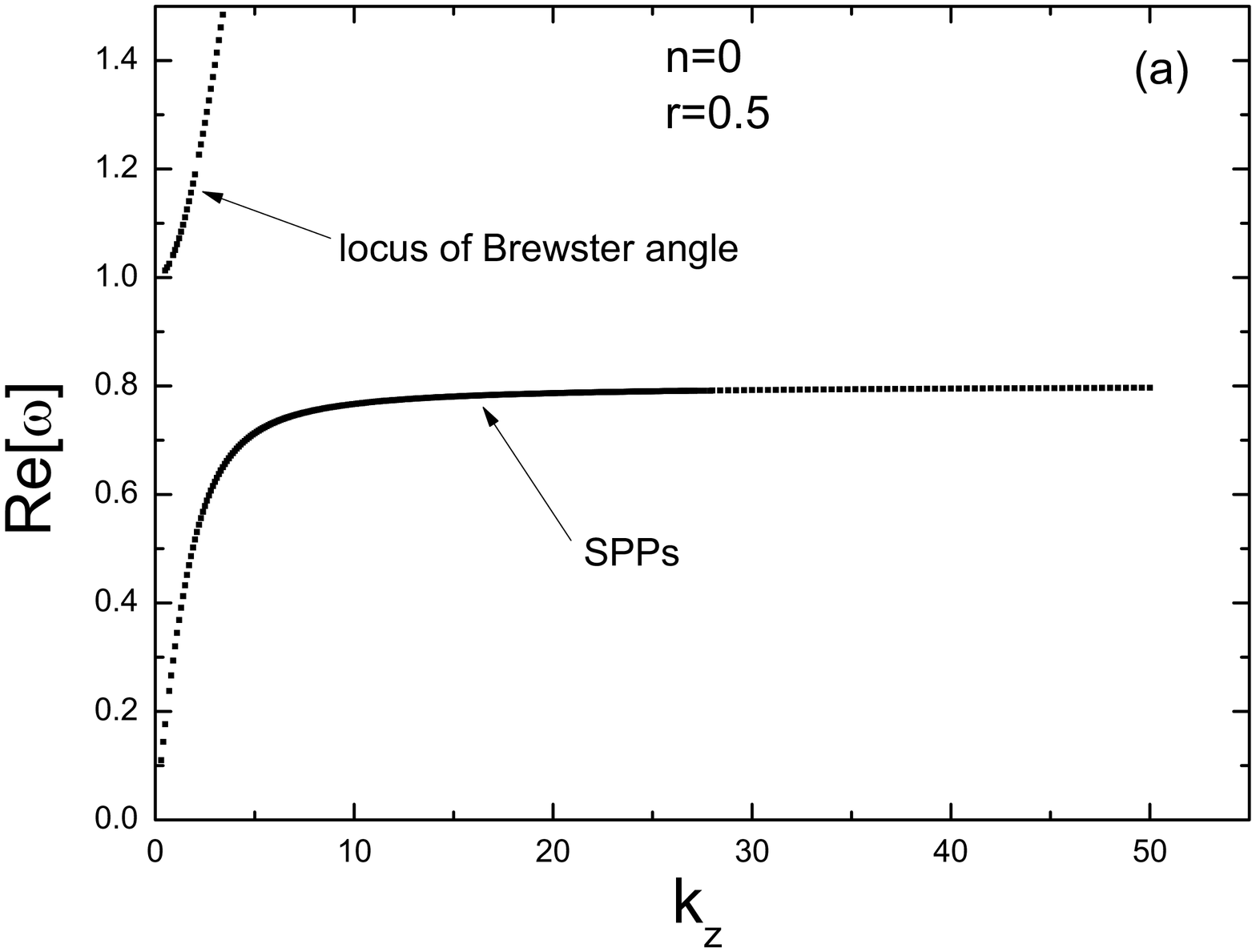} \hspace*{1cm}
\includegraphics[width=2.0in, height=2.7in, angle=270]{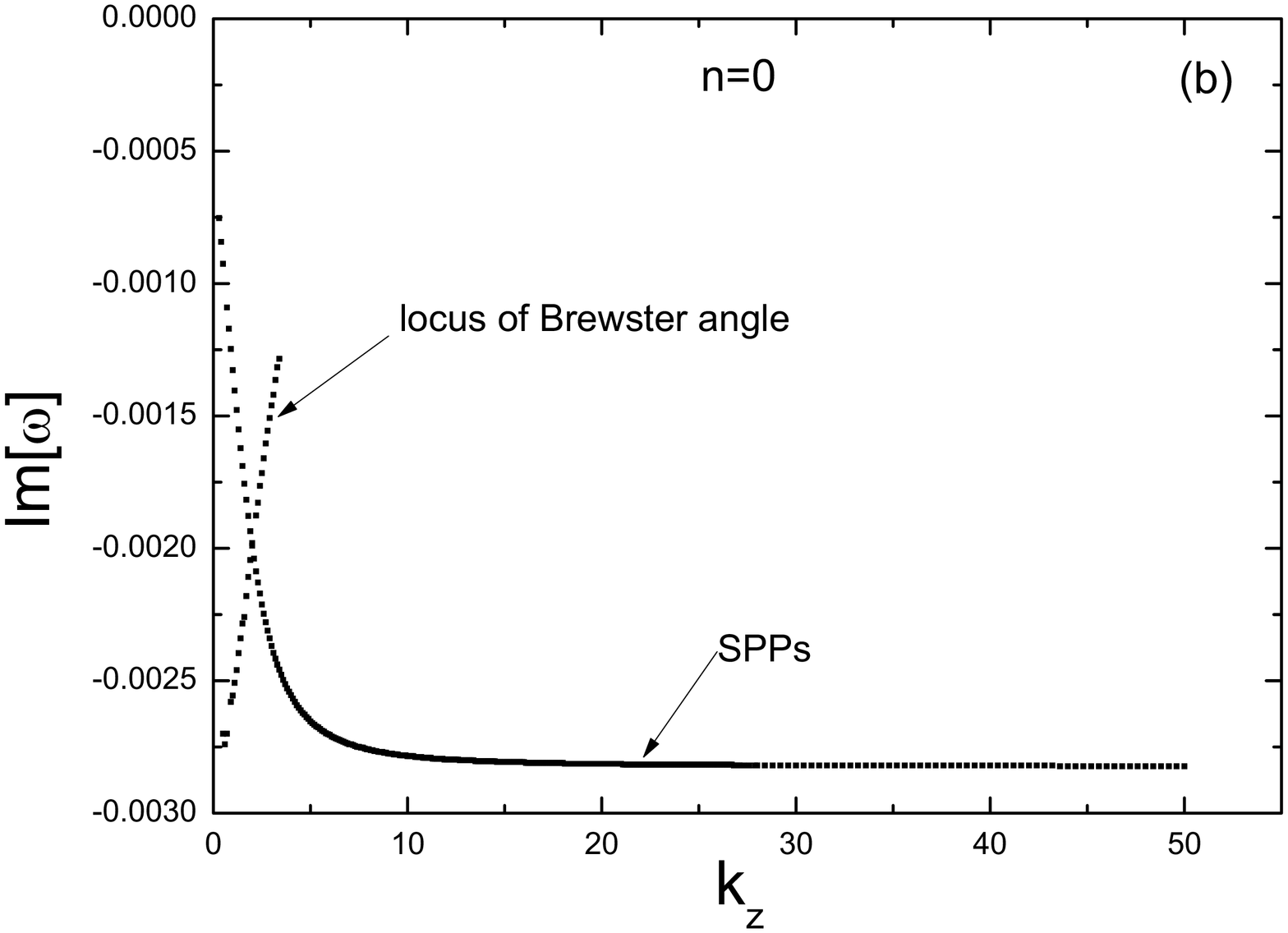}
\\
\includegraphics[width=2.0in, height=2.7in, angle=270]{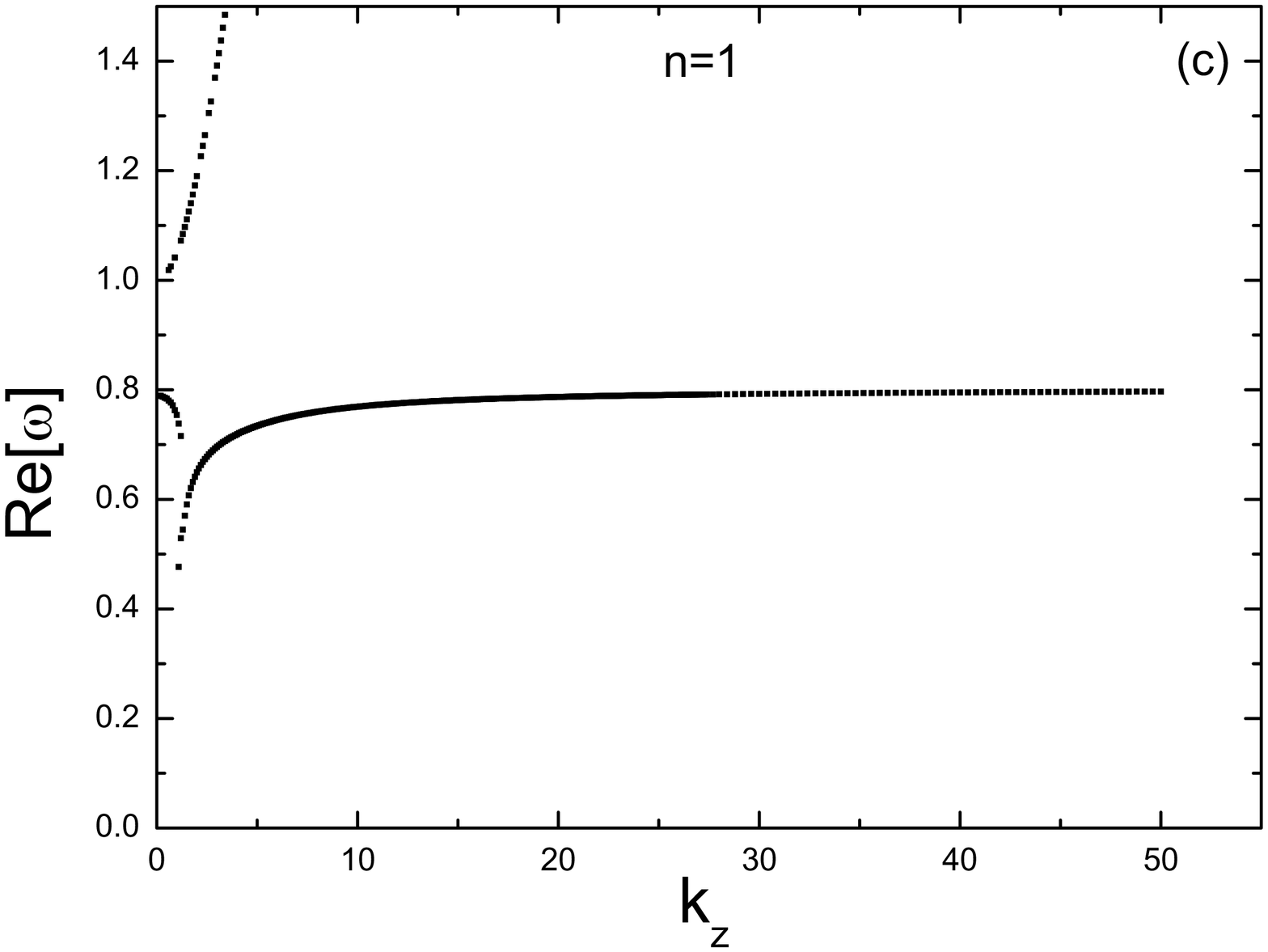}
\hspace*{1cm} \includegraphics[width=2.0in, height=2.7in,
angle=270]{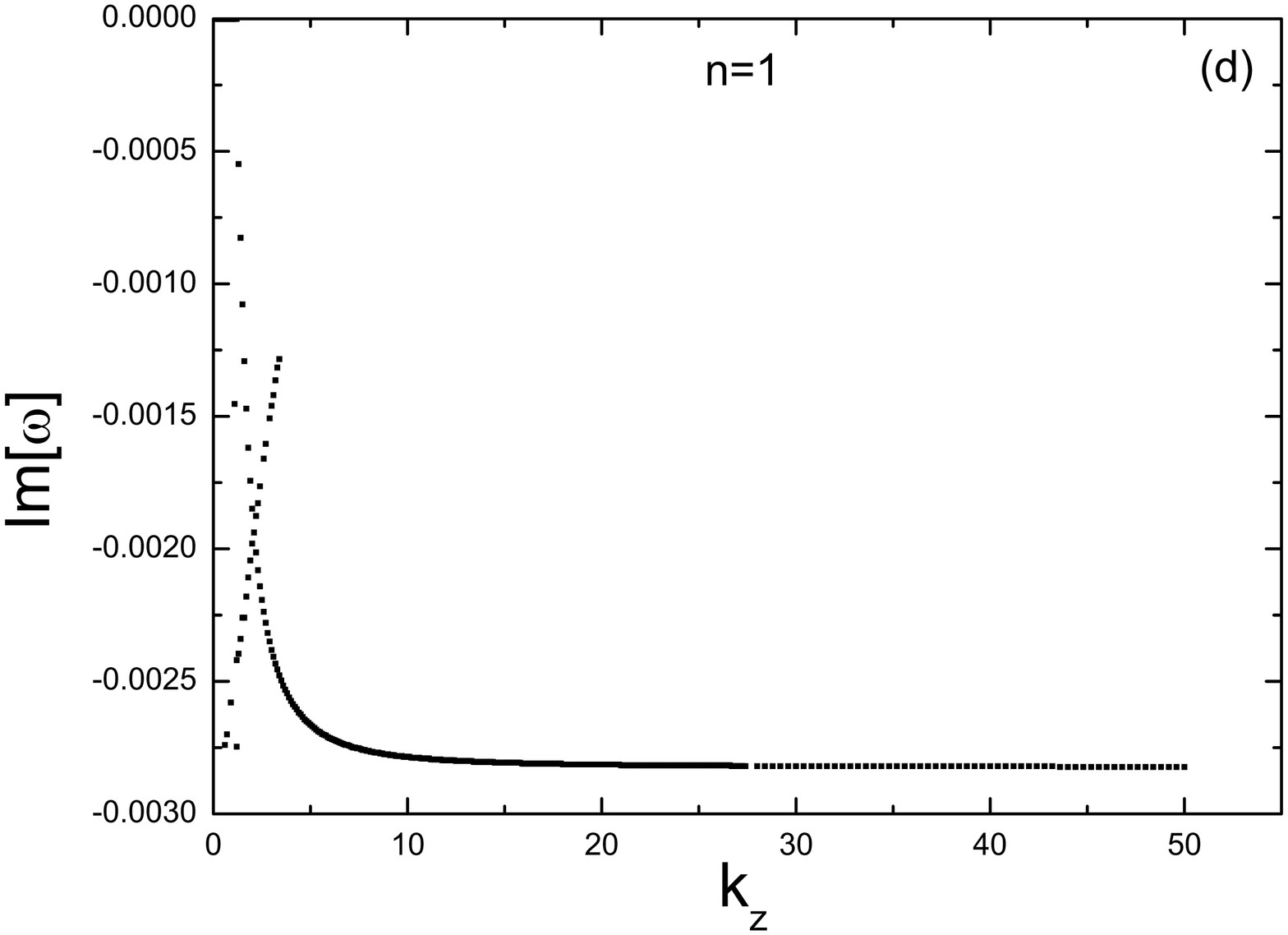}
\\
\includegraphics[width=2.0in, height=2.7in, angle=270]{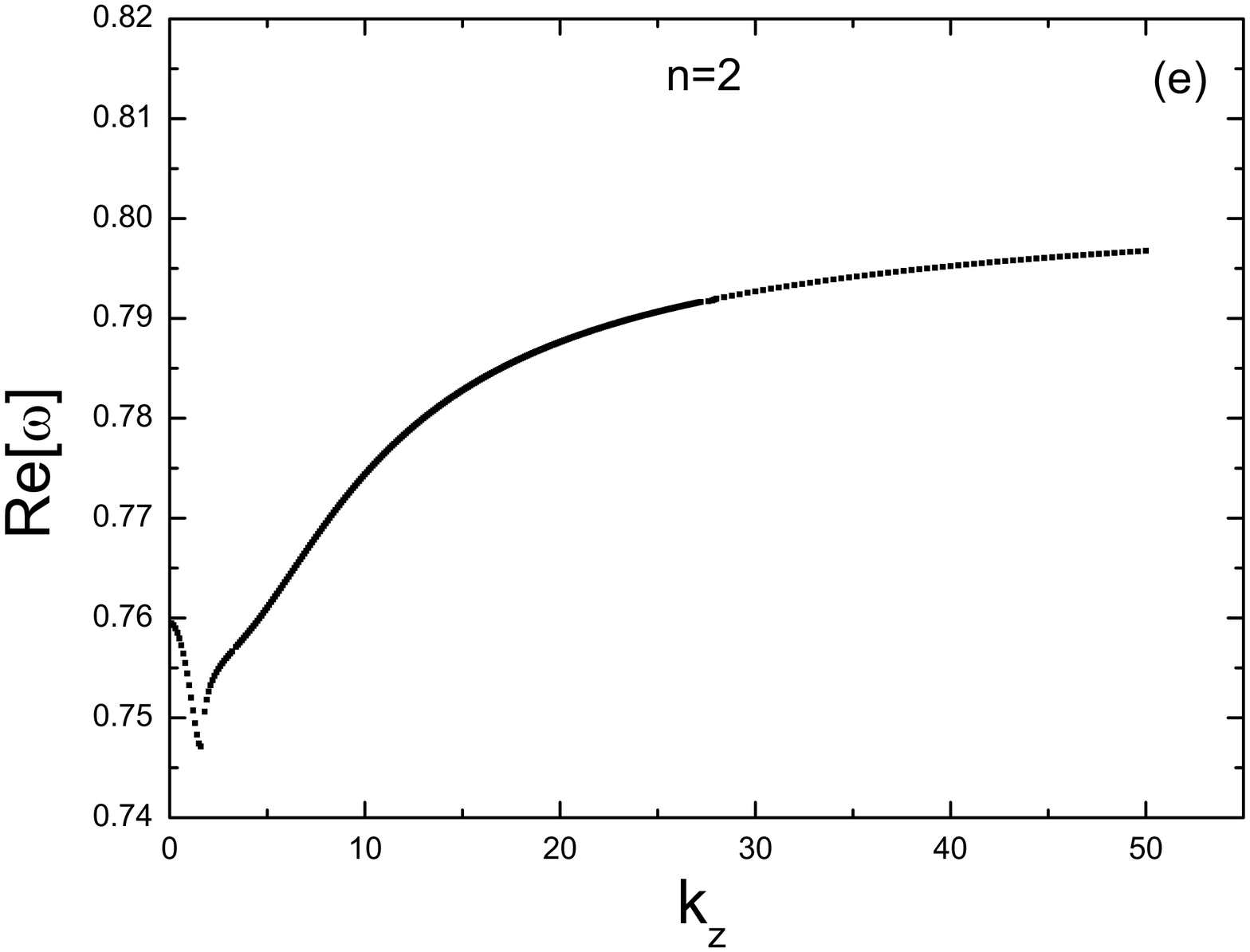}
\hspace*{1cm}
\includegraphics[width=2.0in, height=2.7in, angle=270]{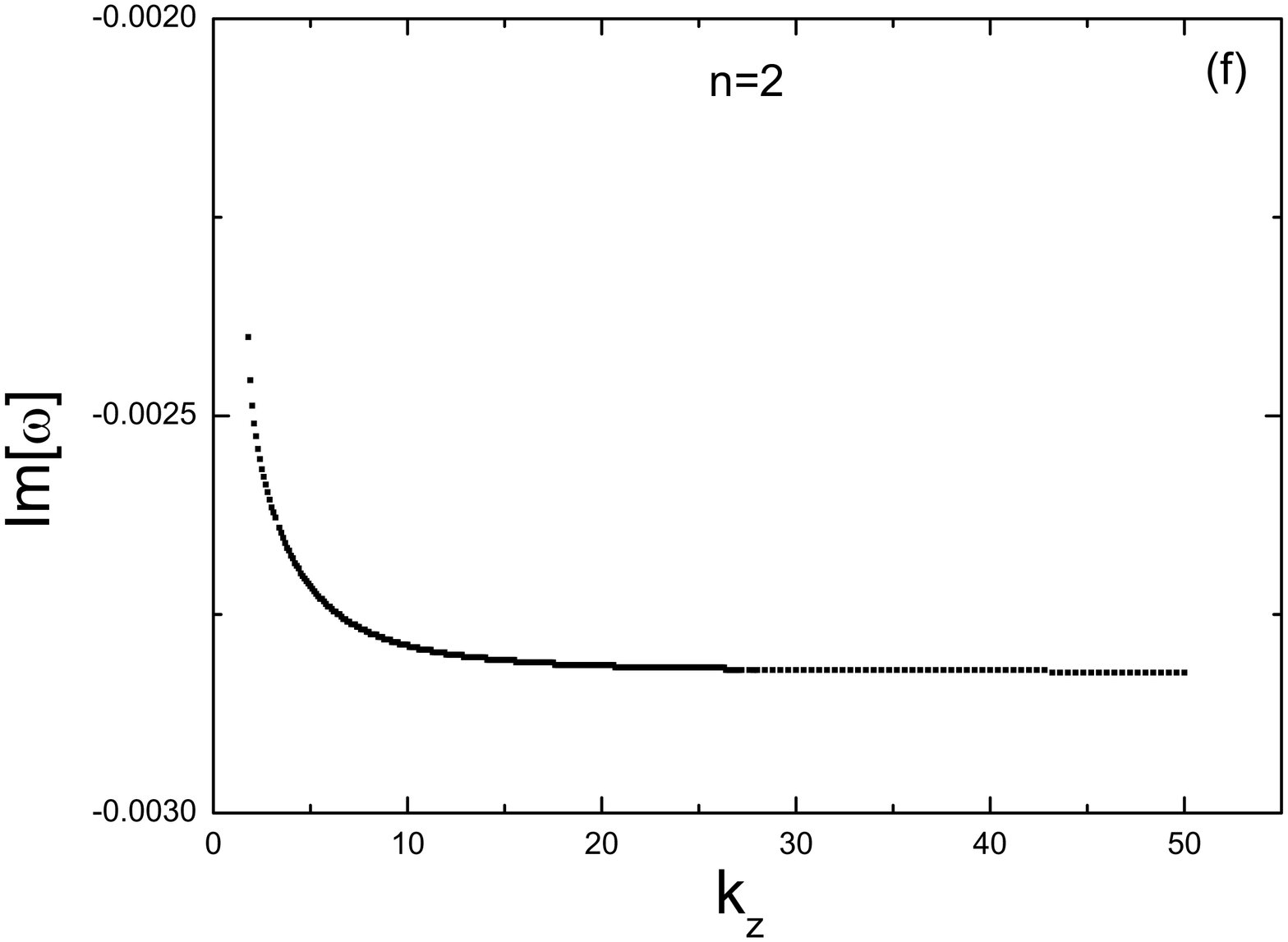}
\\
\includegraphics[width=2.0in, height=2.7in, angle=270]{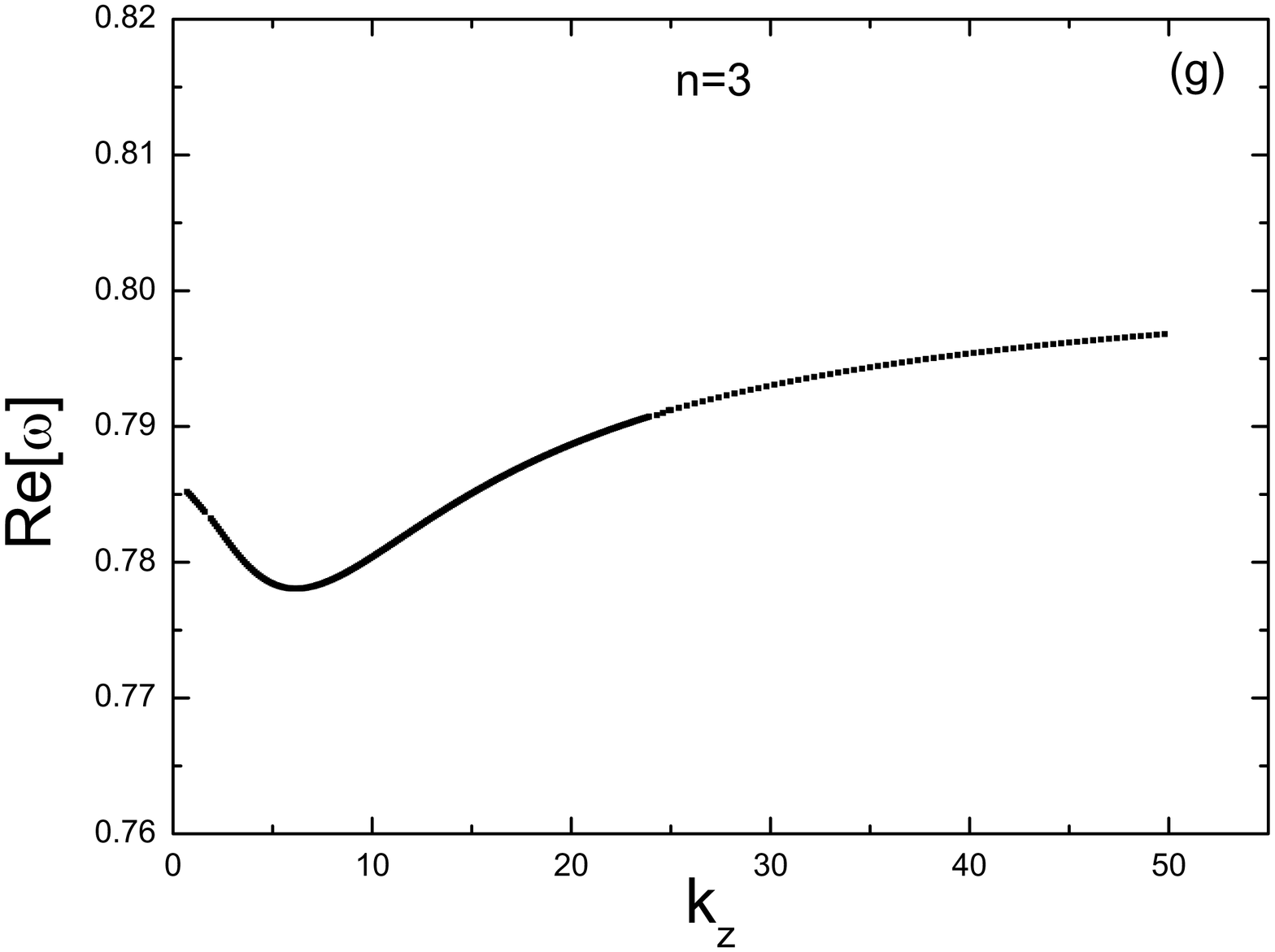}
\hspace*{1cm}\includegraphics[width=2.0in, height=2.7in,
angle=270]{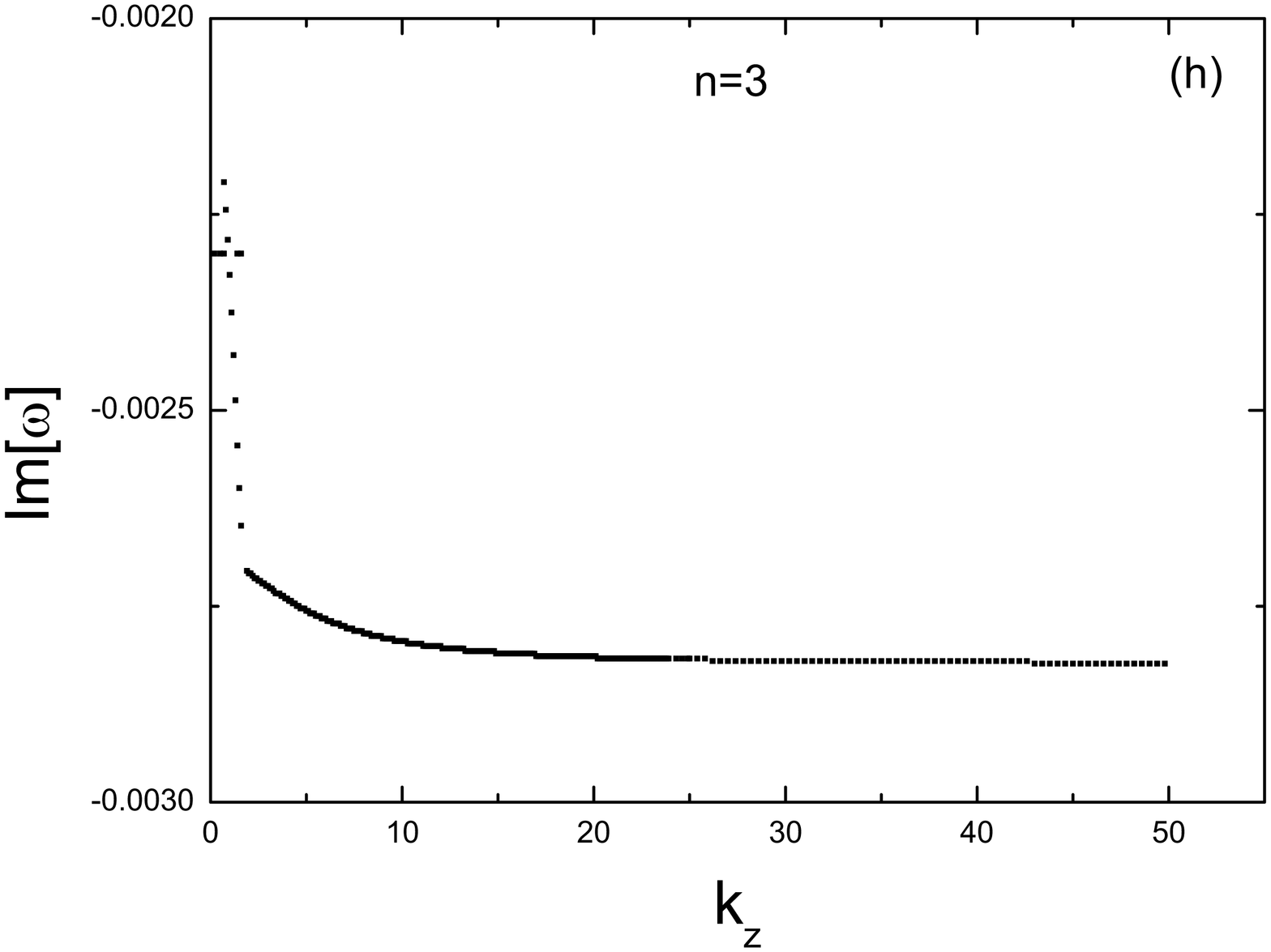}
 \caption{$complex-\omega$
dispersion relations calculated by our method for a cylindrical
metallic nanowire with $\epsilon_0=5.3$, $\epsilon_{\infty}=9.6$ and
$\tau=0.005647$. The renormalized radius $r=0.5$ is used for the
calculation, which is corresponding to the real radius of 26.9$nm$
of the nanowire.} \label{fig.2}
\end{figure*}
Fig.~\ref{fig.2} shows the $complex-\omega$ dispersion relations
calculated from the eq.(\ref{eq.9}) by using our method. For the
comparison between the published results~\cite{Chen1} and our
results, the dielectrics take the values of $\epsilon_0=5.3$ and
$\epsilon_{\infty}=9.6$ with $\tau=0.005647$ in the calculation. The
renormalized radius $r$ takes the value of $r=0.5$ for the nanowire,
which is corresponding to the real radius of 26.9$nm$. Four orders
($n=0,1,2,3$) of the dispersion relations are calculated. The left
column of the figures is for the relation of $Re[\omega]$ and $k_z$
and the right column is for the relation of $Im[\omega]$ and $k_z$.
In the calculation, the mesh size is 0.01 and the error allowance
$\delta$ is set to be $\delta=0.1$. It is shown that there exist two
dispersion branches in the figures of $n=0$ and $n=1$, shown in
figs.~\ref{fig.2}(a)-(d). One branch is an asymptotic curve with the
frequency approaching the surface plasmon frequency
$\omega_{sp}=\sqrt{\epsilon_{\infty}/(\epsilon_0+\epsilon_{\infty})}\approx
0.8$, which represents the SPPs surface wave. Another branch having
the frequency above 1 is not a surface wave but identified as the
locus of the Brewster angle.~\cite{Archambault1} In our calculation,
the SPPs dispersion curve are found by the using of the criterion
(\ref{eq.4}) while the locus of the Brewster angle needs the
criterion (\ref{eq.8}), reflecting that the relation between the
$k_z$ and $Re[\omega]$ behaves parabola-like when it is close to the
locus of the Brewster angle. For the order $n>1$, the relation
between the $\omega$ and $k_z$ is nearly dispersionless and no locus
of the Brewster angle can be observed anymore. The imaginary parts
of the frequencies are very small, which have been shown in the
right figures. Our calculated results are coincident to the
published results,~\cite{Chen1} confirming the validity of our
method. Our results exhibit that two dispersion branches can be
obtained in once calculation by our method, however, which can not
be achieved by the NP method if the frequency range includes the two
curves.

\subsection{$complex-k$ solution}
\begin{figure*}[t!]
\centering
\includegraphics[width=2.0in, height=2.7in,
angle=270]{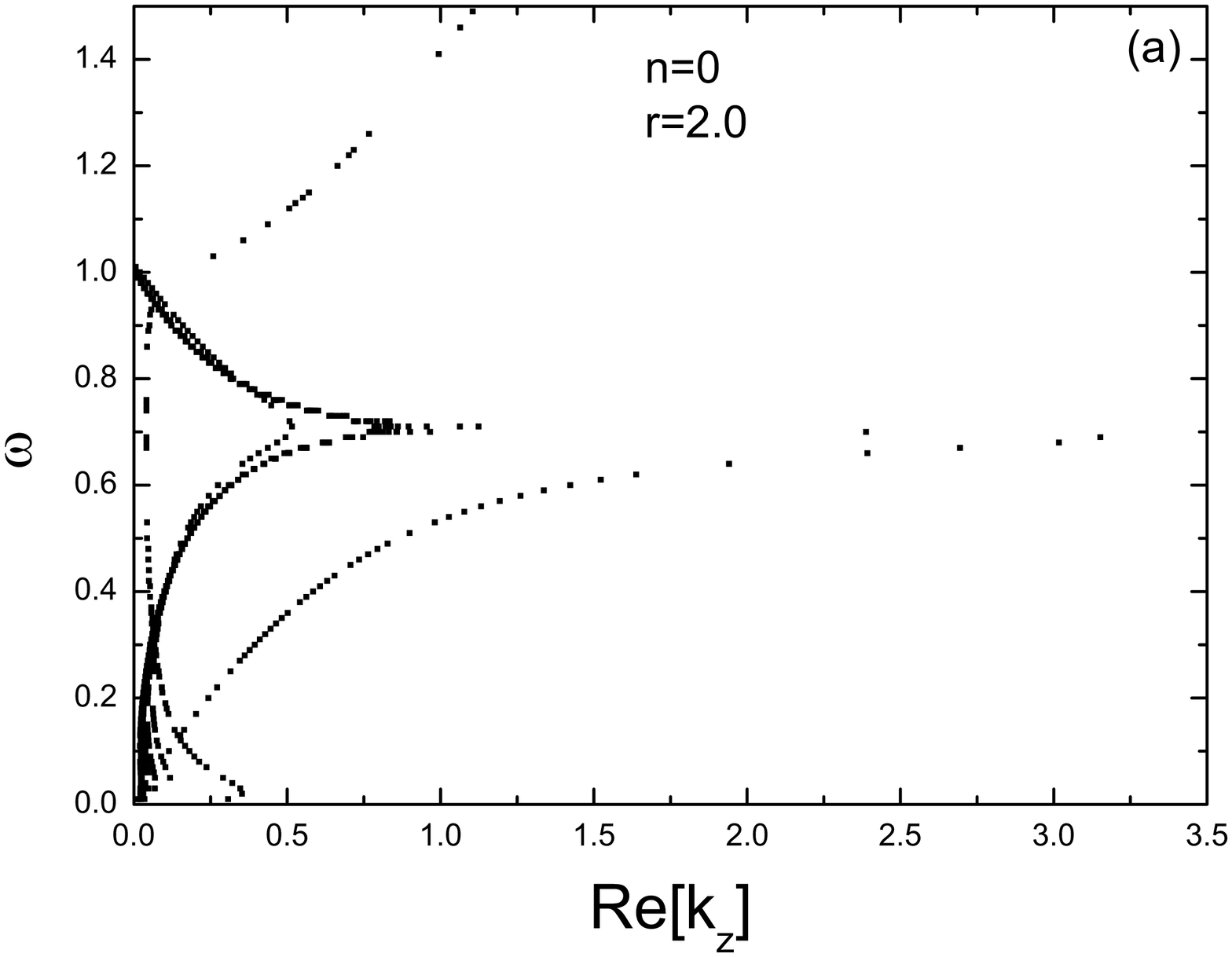}\hspace*{1cm}
\includegraphics[width=2.0in, height=2.7in, angle=270]{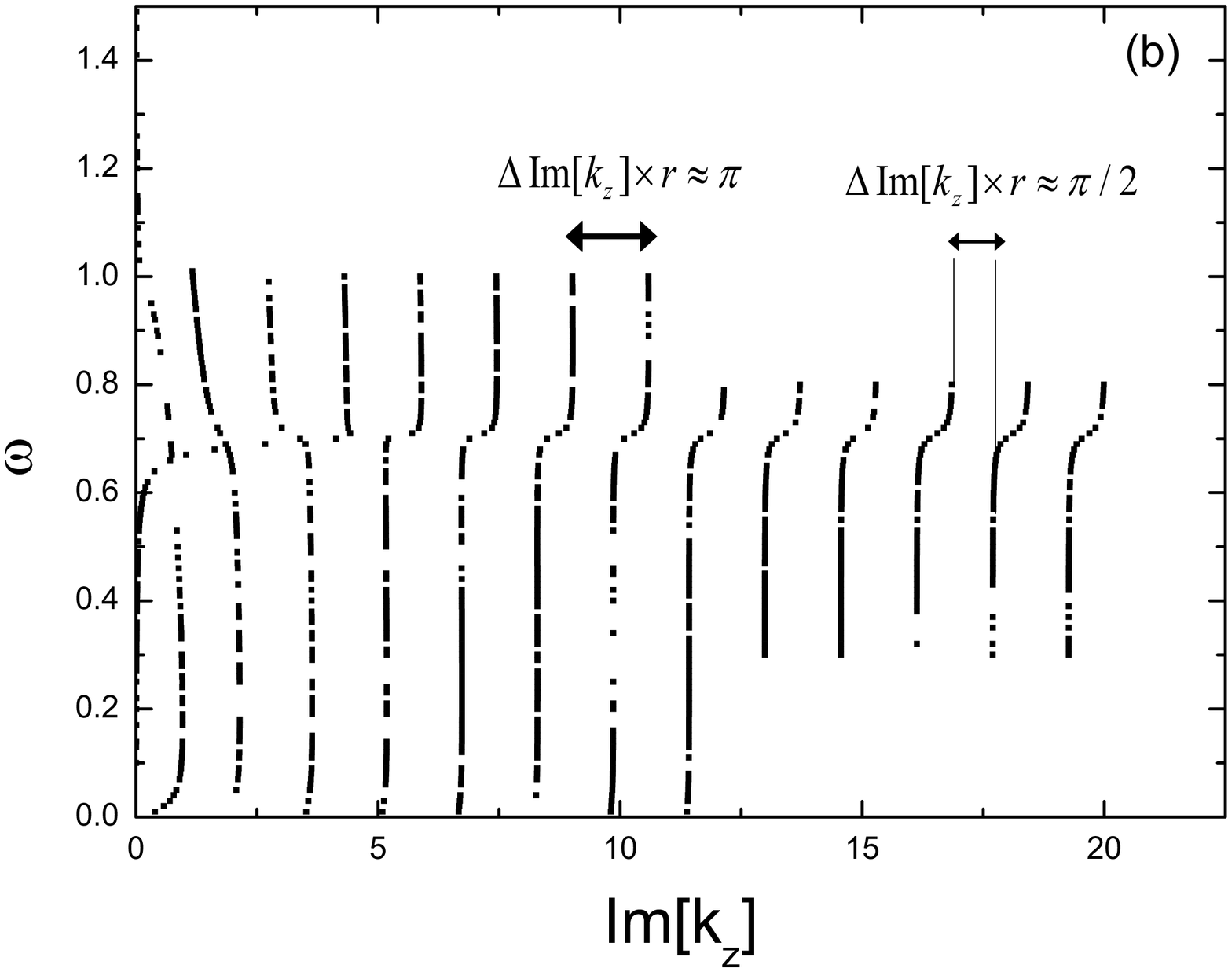}
\\
\includegraphics[width=2.0in, height=2.7in,
angle=270]{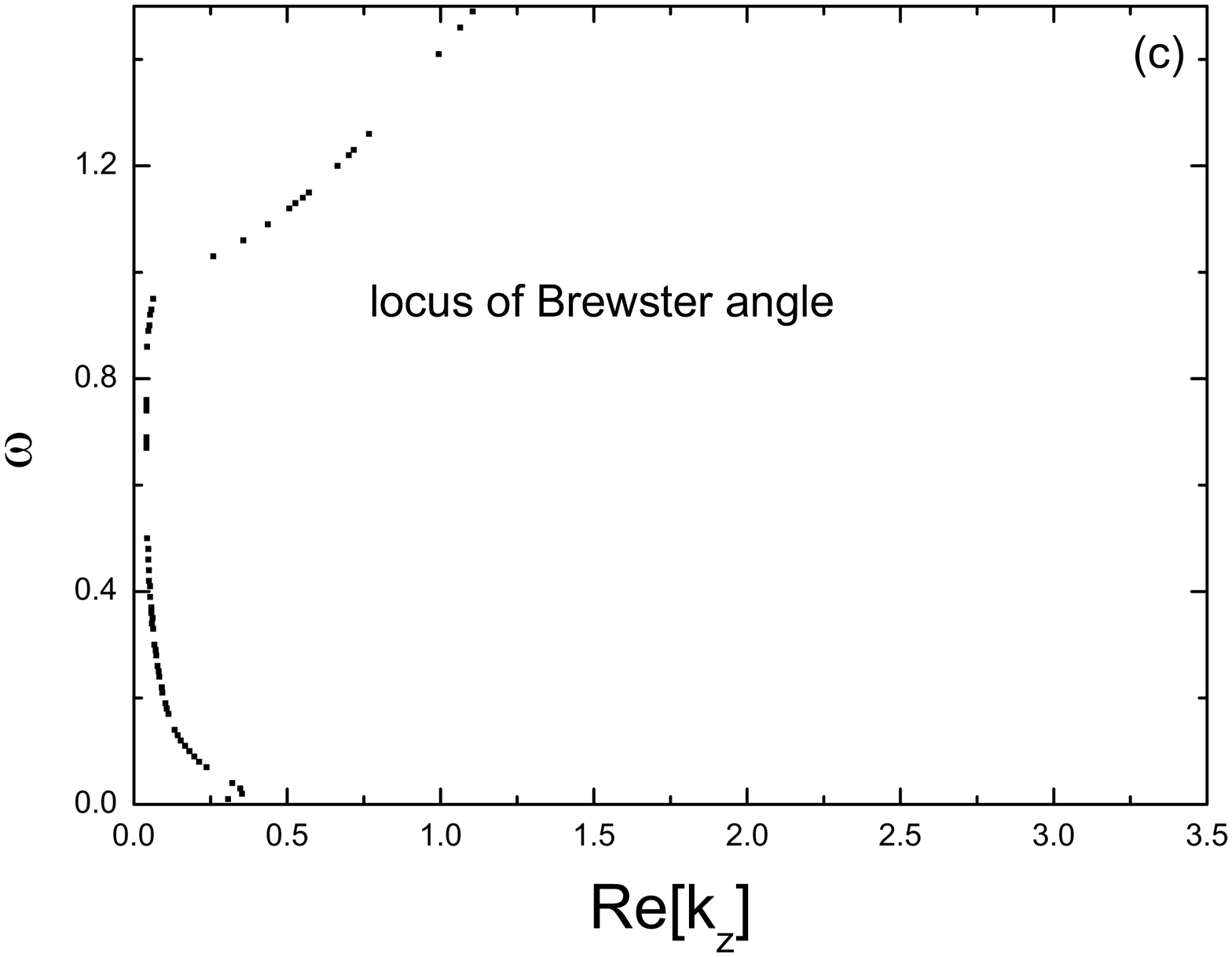} \hspace*{1cm}
\includegraphics[width=2.0in, height=2.7in, angle=270]{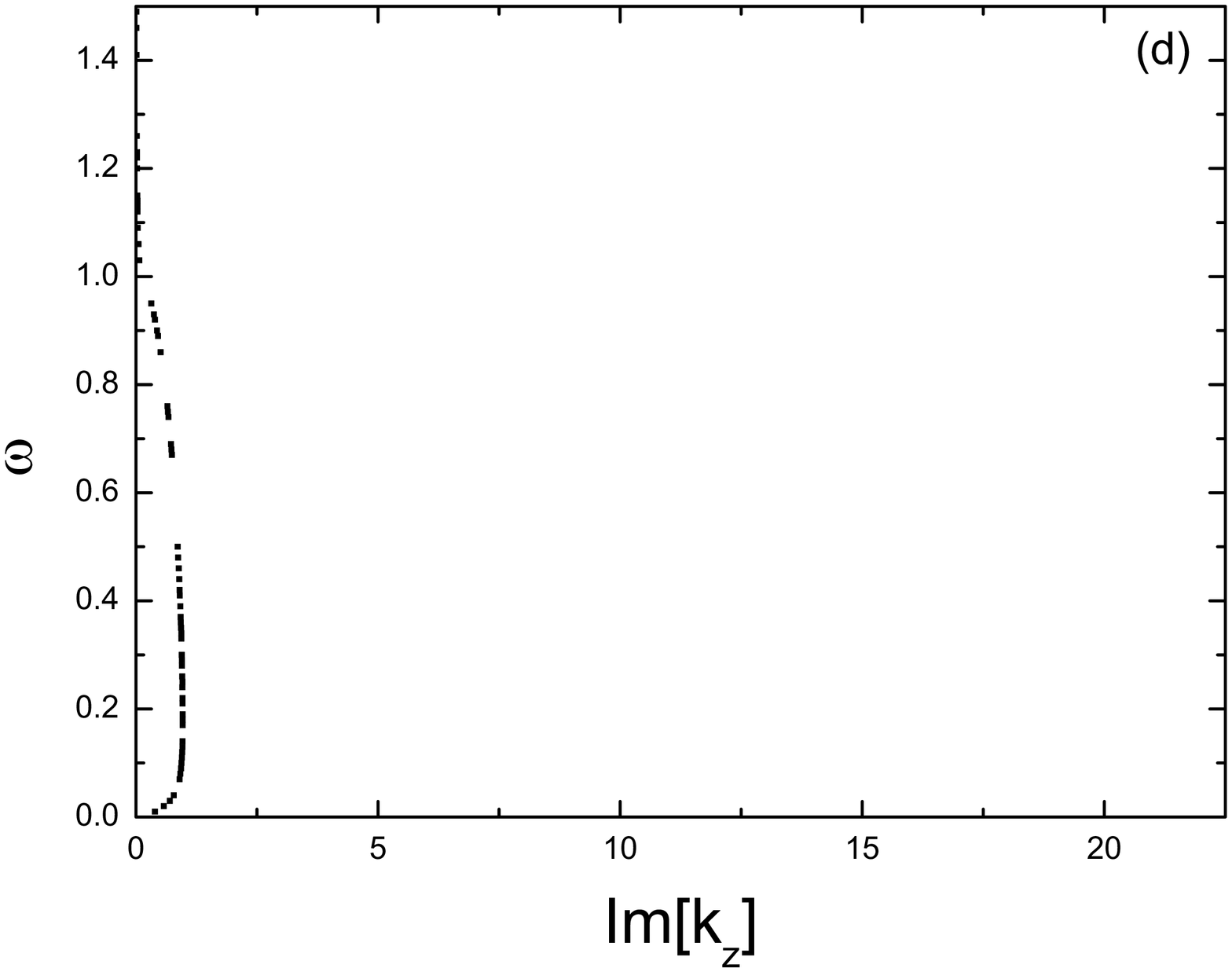}
\\
\includegraphics[width=2.0in, height=2.7in,
angle=270]{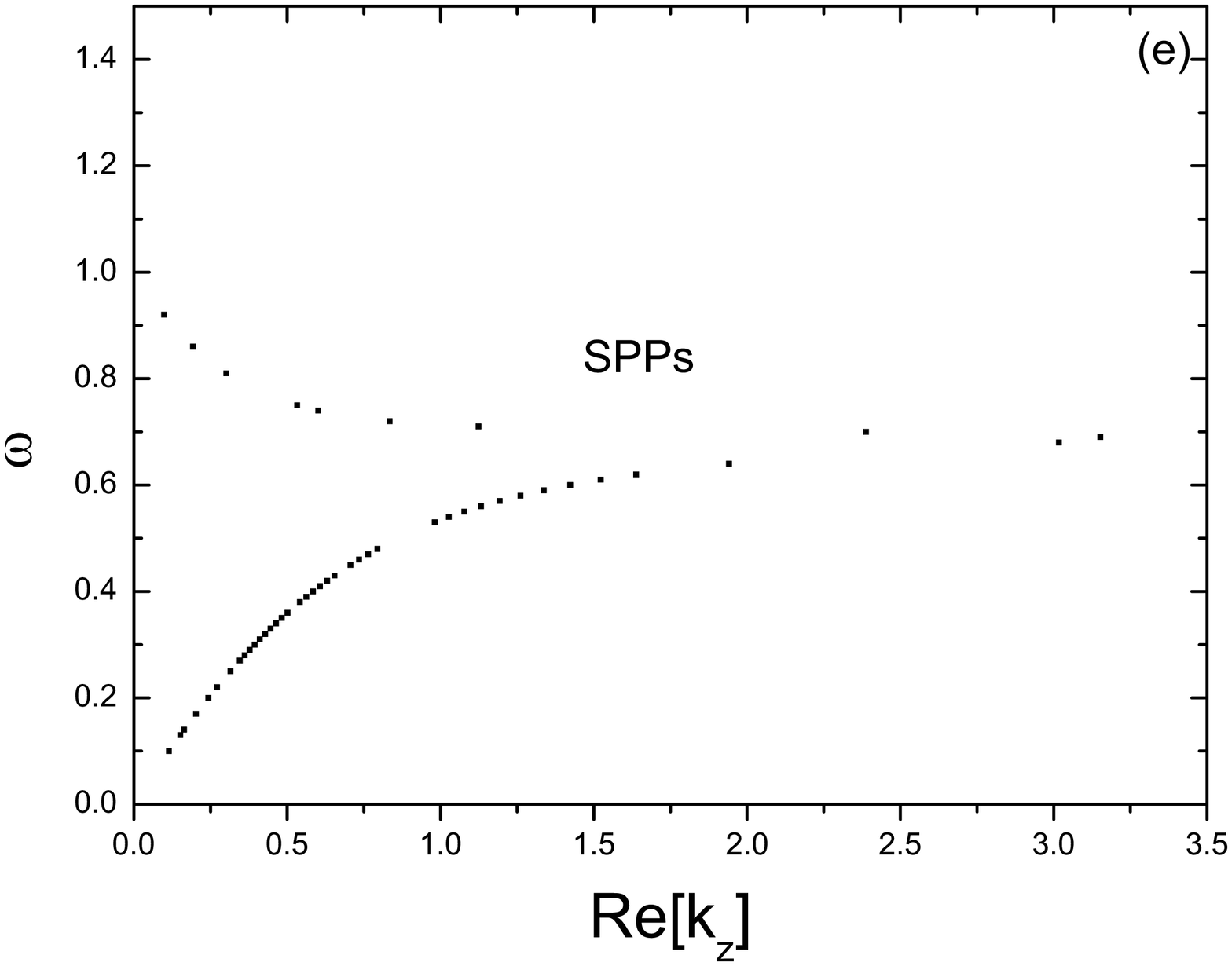} \hspace*{1cm}
\includegraphics[width=2.0in, height=2.7in, angle=270]{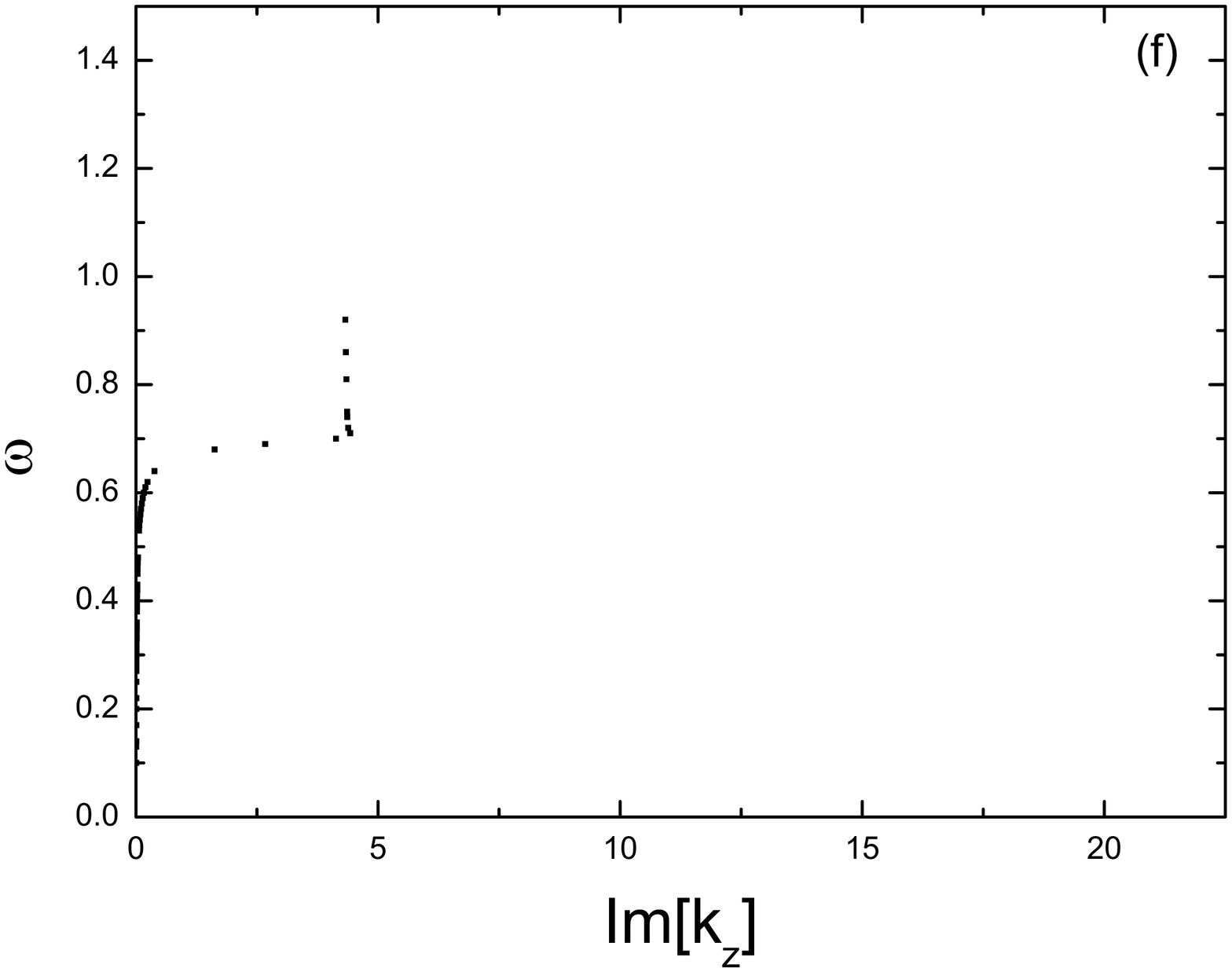}
\\
\includegraphics[width=2.0in, height=2.7in,
angle=270]{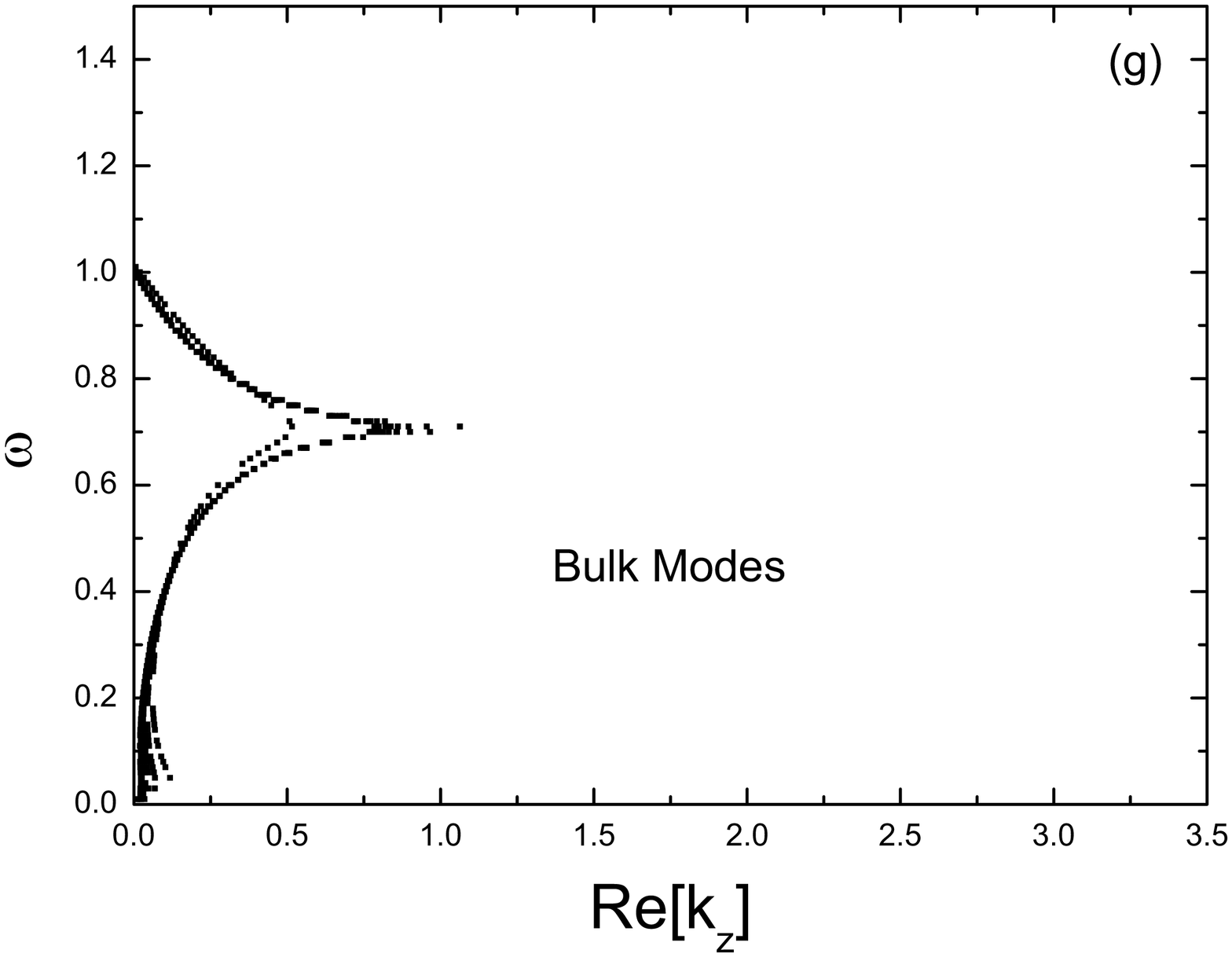} \hspace*{1cm}
\includegraphics[width=2.0in, height=2.7in, angle=270]{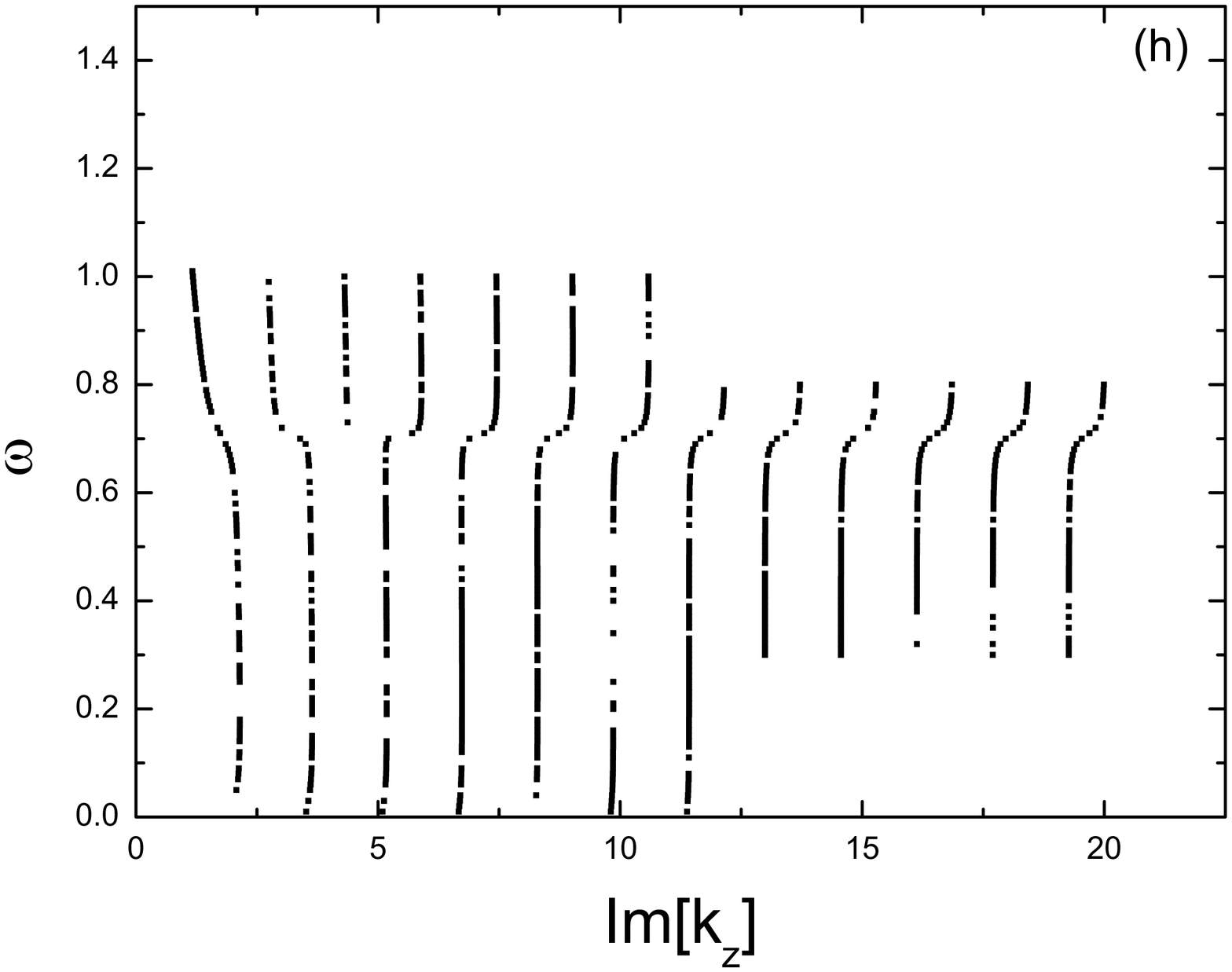}
\caption{$complex-k$ dispersion relations calculated by our method
for a cylindrical metallic nanowire with $\epsilon_0=1$,
$\epsilon_{\infty}=1$ and $\tau=0.039062$. In the calculation, $r=2$
and $n=0$ are used. The renormalized radius $r=2$ is corresponding
to the real radius of 105$nm$ of the nanowire.} \label{fig.3}
\end{figure*}
We have repeated the reported results~\cite{Novotnv1} by using our
method with $\epsilon_0=1$, $\epsilon_{\infty}=1$, $\tau=0.039062$
and $n=1$. However, we only show our result with $n=0$ in
fig.~\ref{fig.3} since in this case an approximate analytical
solution can be derived to verify our calculated results. The
renormalized radius of the nanowire for the calculation is $r=2$,
which is corresponding to the real radius of
105$nm$.~\cite{Novotnv1} By using our method, full solutions to the
complex dispersion equation can be obtained, exhibiting many modes
in fig.~\ref{fig.3}(a) and (b). Those modes can be classified into
three, which have been shown in fig.~\ref{fig.3}(c)-(h). Each figure
in the left side responses for the relation of $\omega$ and
$Re[k_z]$, and the corresponding figure at its right side is for the
relation of $\omega$ and $Im[k_z]$. Dispersion curves in
fig.~\ref{fig.3}(c) and (d) exhibit the locus of the Brewster
angle.~\cite{Archambault1} Fig.~\ref{fig.3}(e) and (f) represent the
SPPs mode with a back bending in the curve. The back bending is
induced by the metal loss.~\cite{Wan1} For a nanowire fabricated
with perfect metal, the SPPs mode then is an asymptotic curve to
infinity when the frequency approaches the $\omega_{sp}=1/\sqrt{2}$.
The third class of the modes in fig.~\ref{fig.3}(g) and (h) have an
infinite mode number and the $Re[k_z]$ of those modes are small
compared to $Im[k_z]$. Those modes have been defined as bulk modes
(BMDs),~\cite{Novotnv1} but never been reported in this case.
Therefore, there exists one question that if the BMDs are wrong
solutions in our calculation. To answer this question, we derive an
approximate solution to the complex dispersion equation in the
following to verify our method.
\begin{figure*}[t!]
\centering
\includegraphics[width=2.0in, height=2.7in, angle=270]{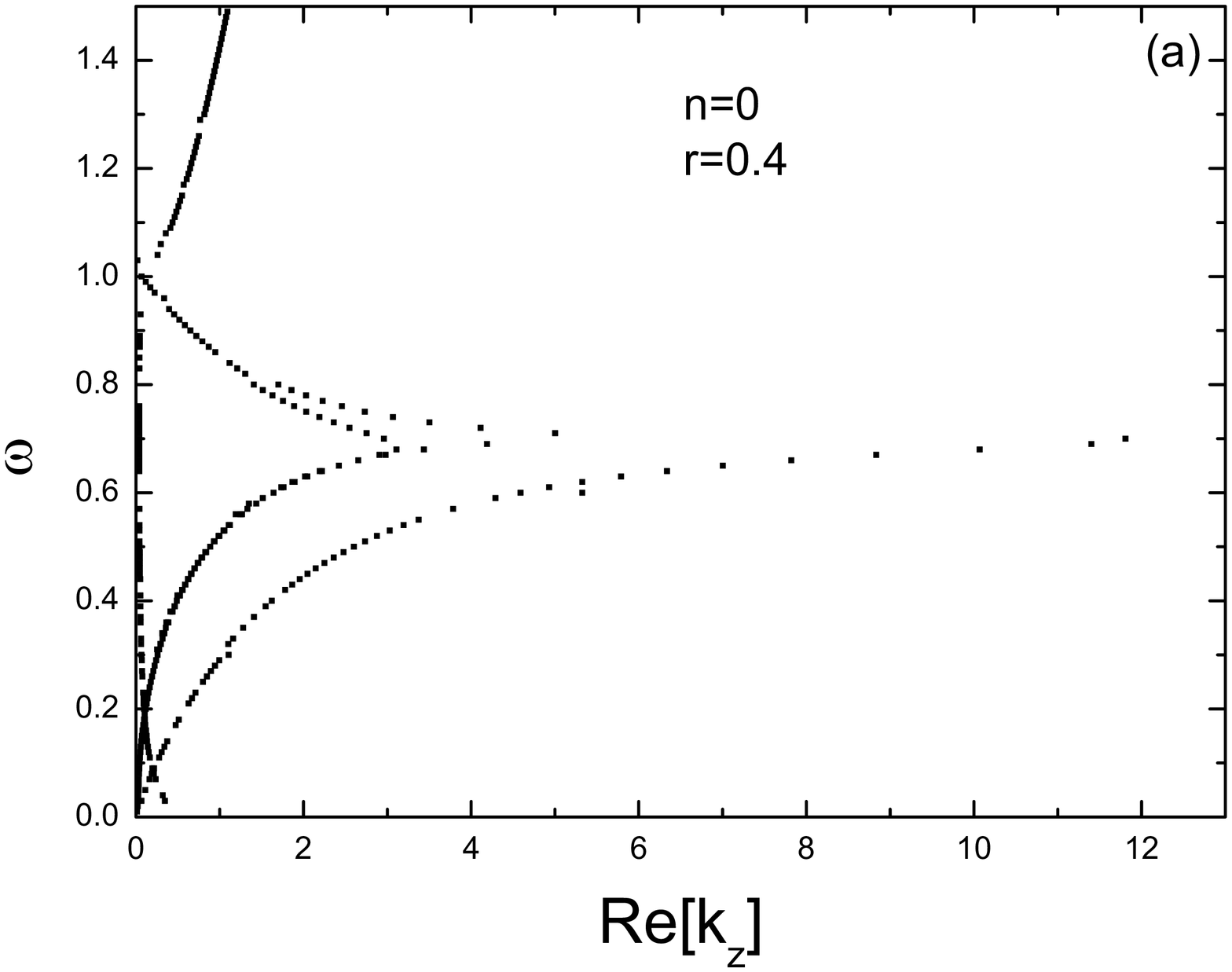}
\hspace*{1cm}
\includegraphics[width=2.0in, height=2.7in, angle=270]{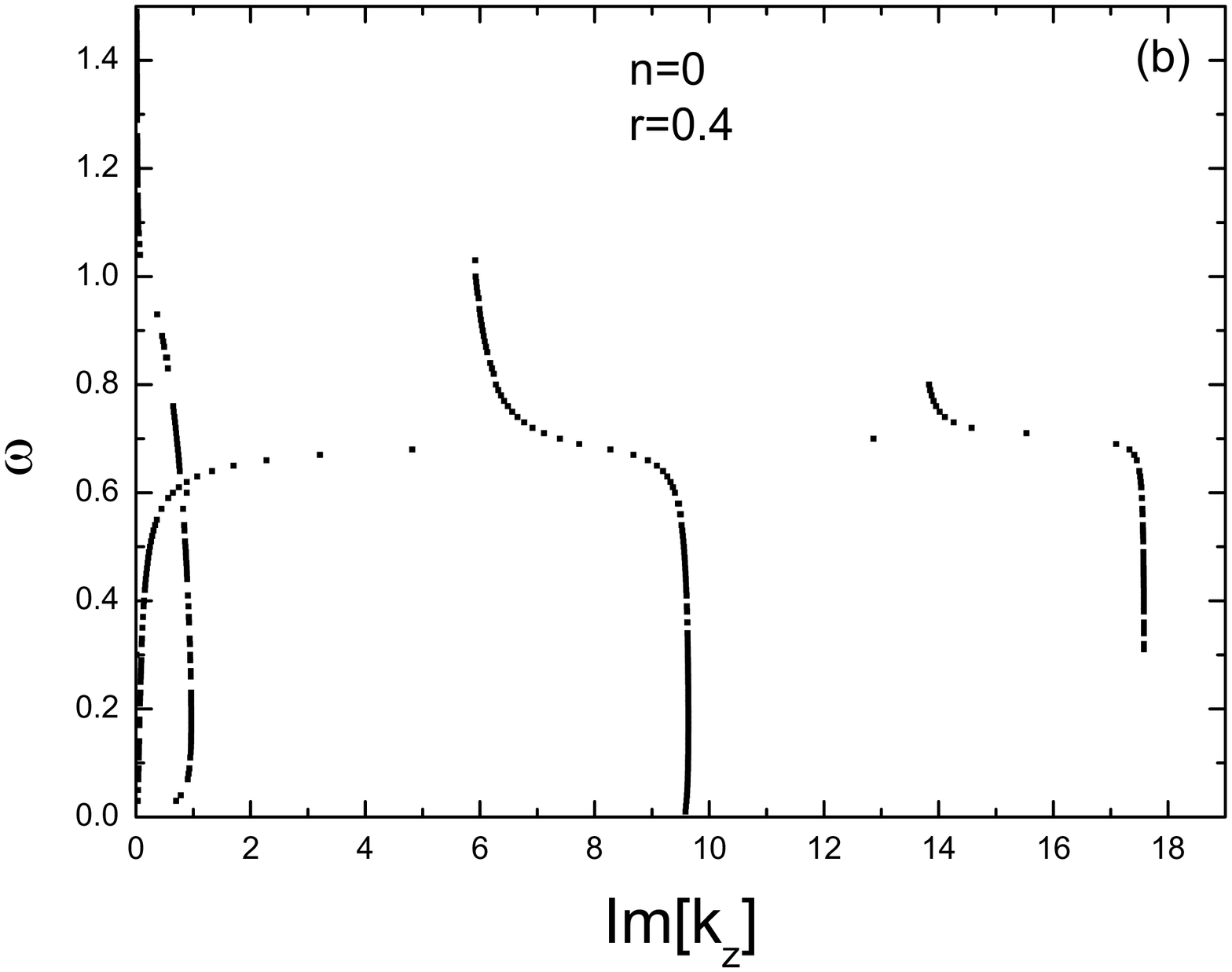}
\\
\includegraphics[width=2.0in, height=2.7in, angle=270]{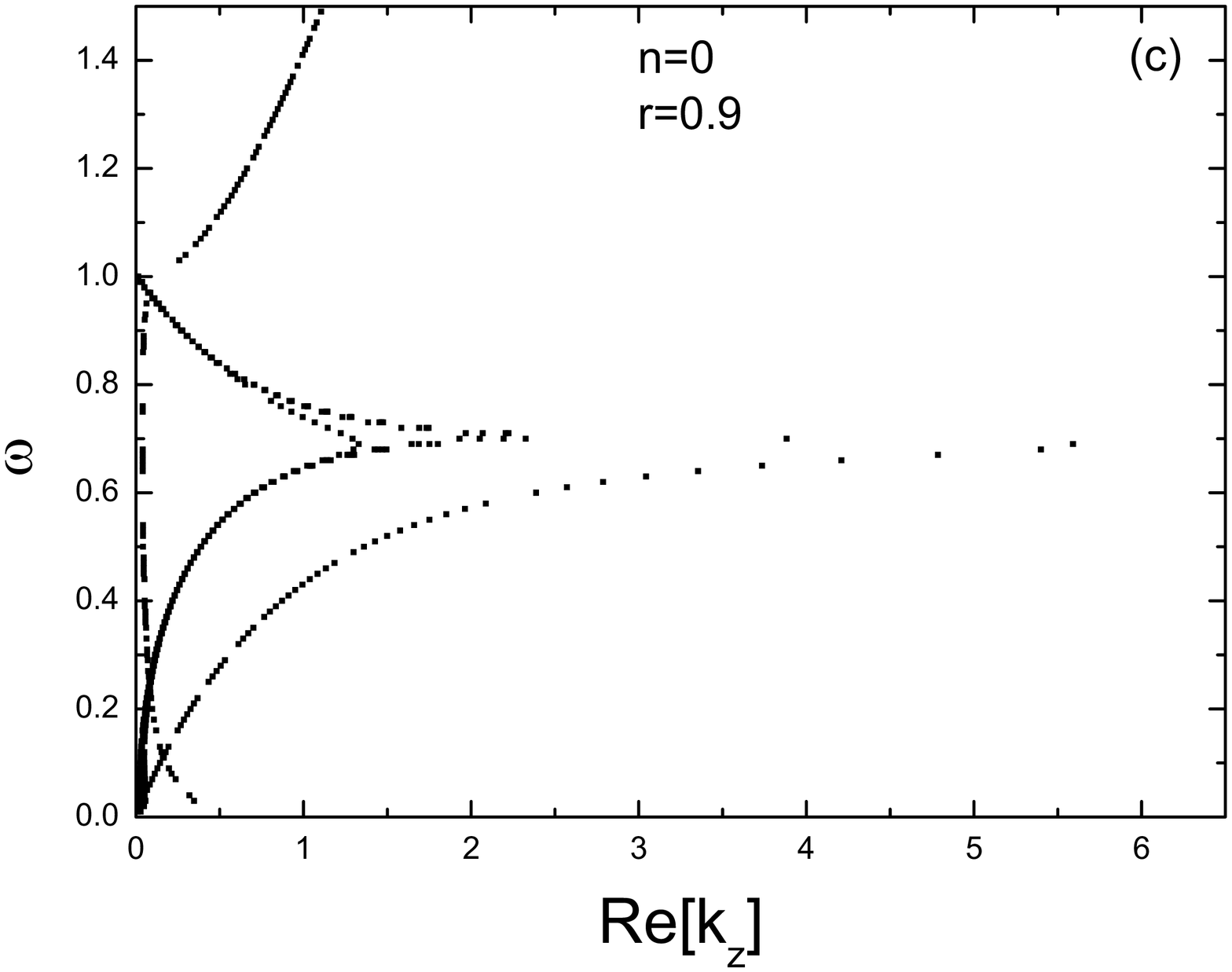}
\hspace*{1cm}
\includegraphics[width=2.0in, height=2.7in, angle=270]{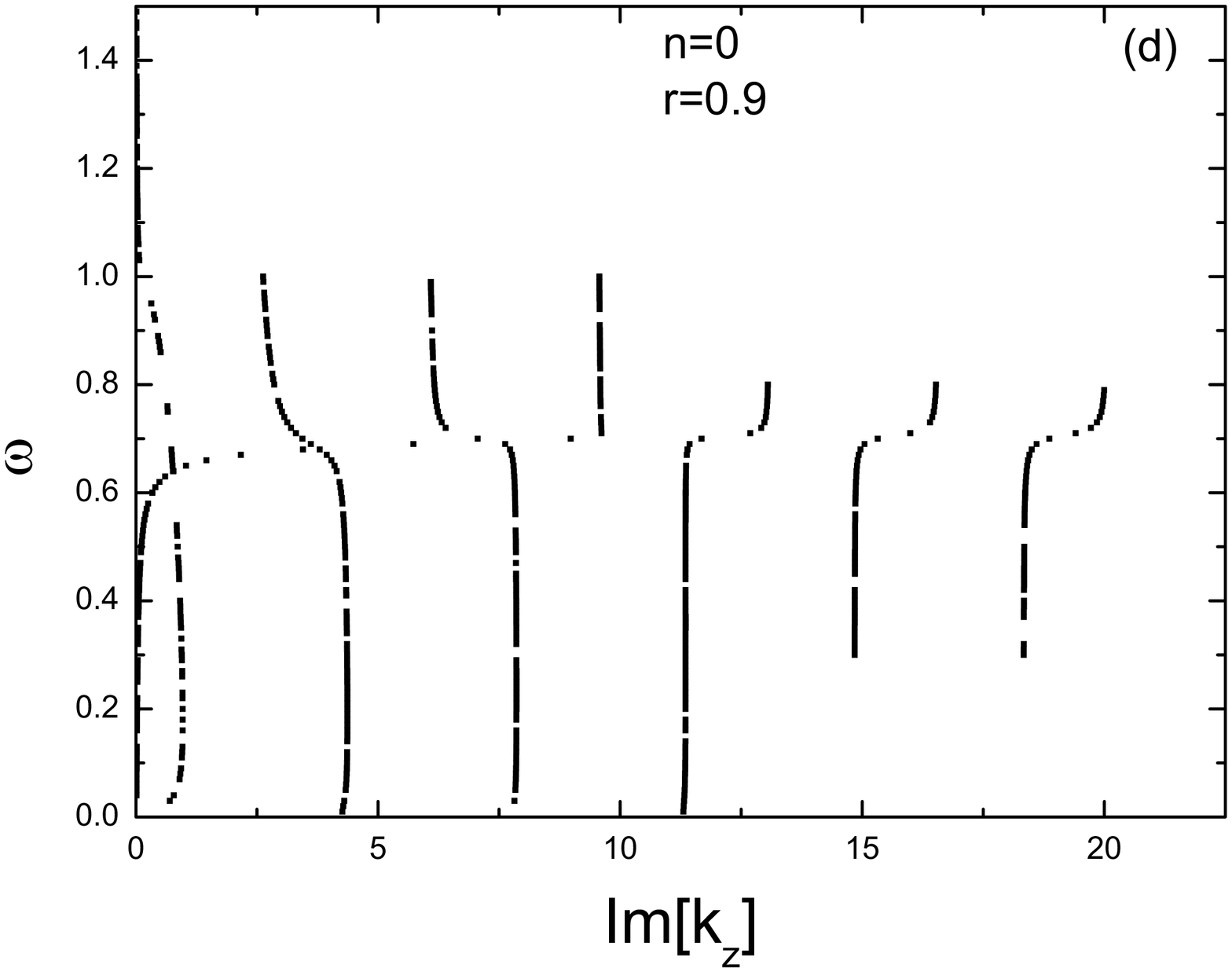}
\\
\includegraphics[width=2.0in, height=2.7in, angle=270]{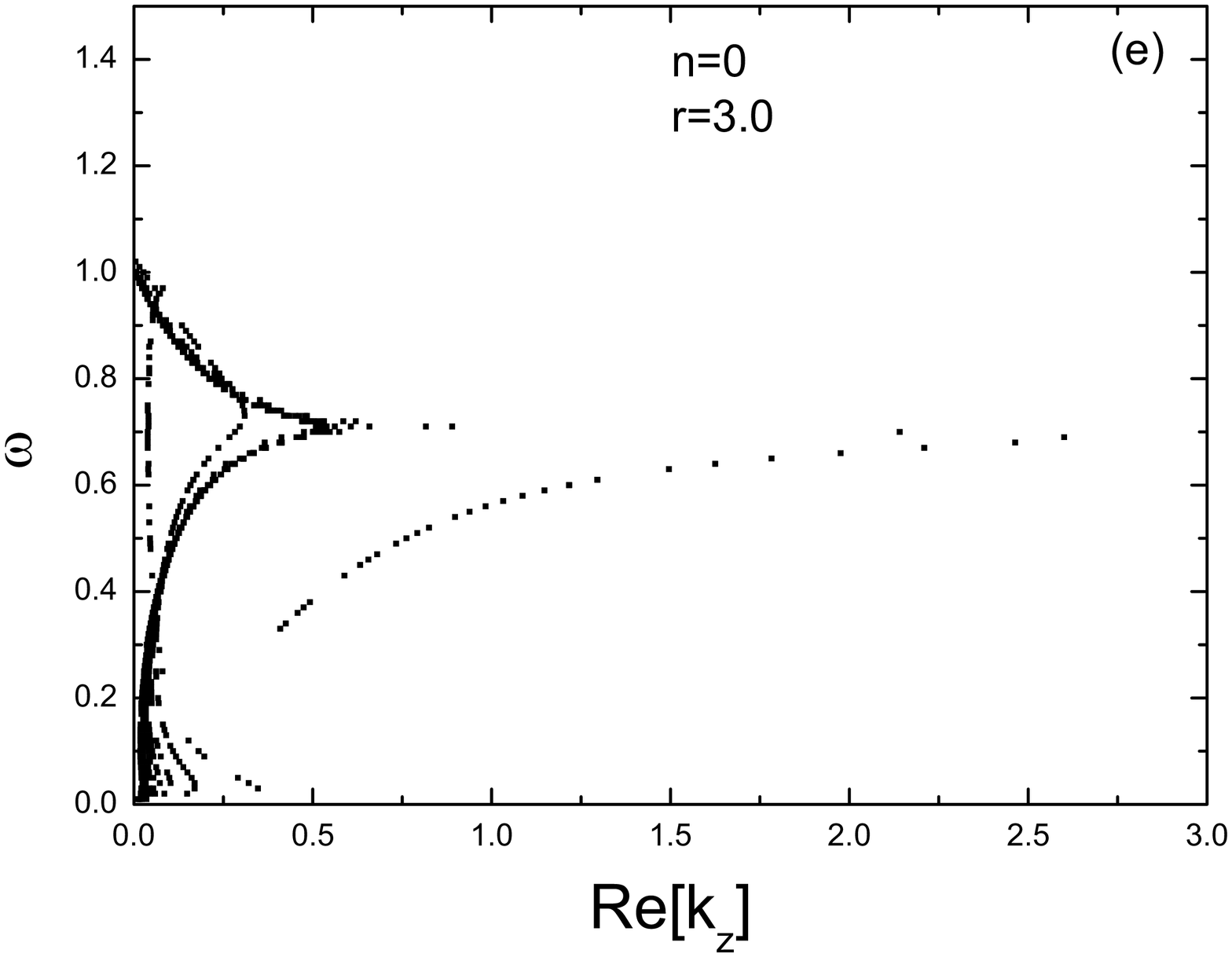}
 \hspace*{1cm}
\includegraphics[width=2.0in, height=2.7in, angle=270]{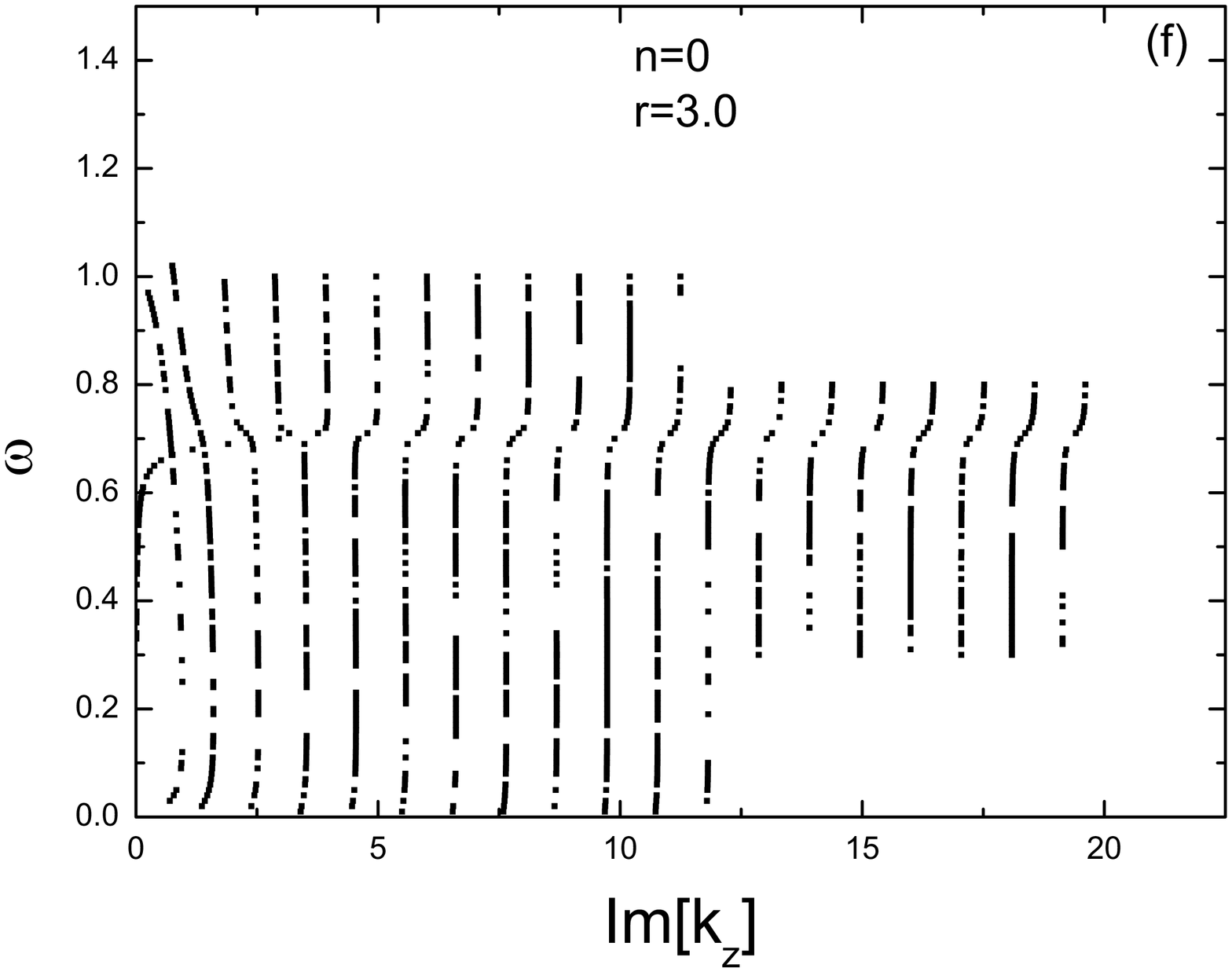}
\\
\includegraphics[width=2.0in, height=2.7in, angle=270]{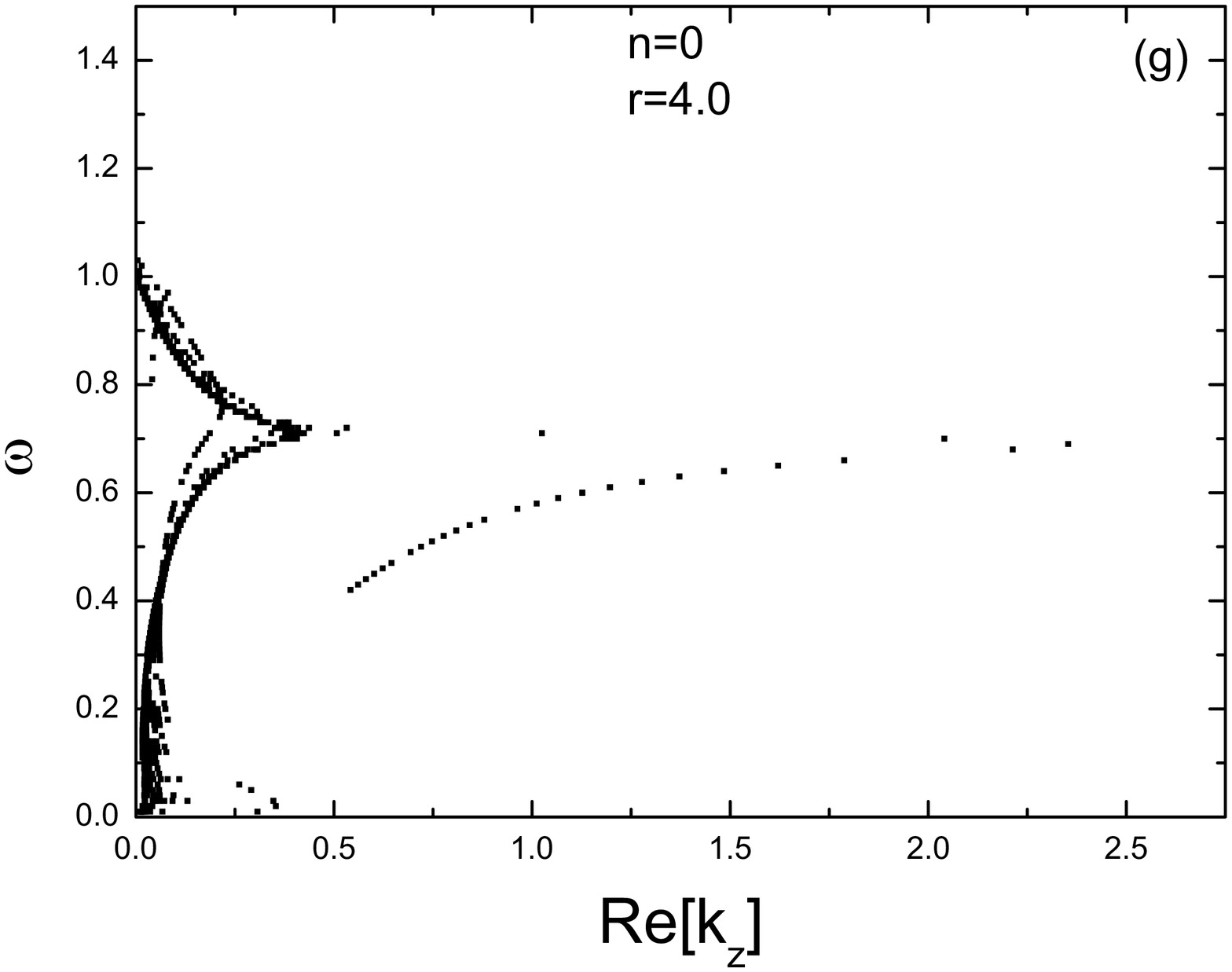}
\hspace*{1cm}
\includegraphics[width=2.0in, height=2.7in, angle=270]{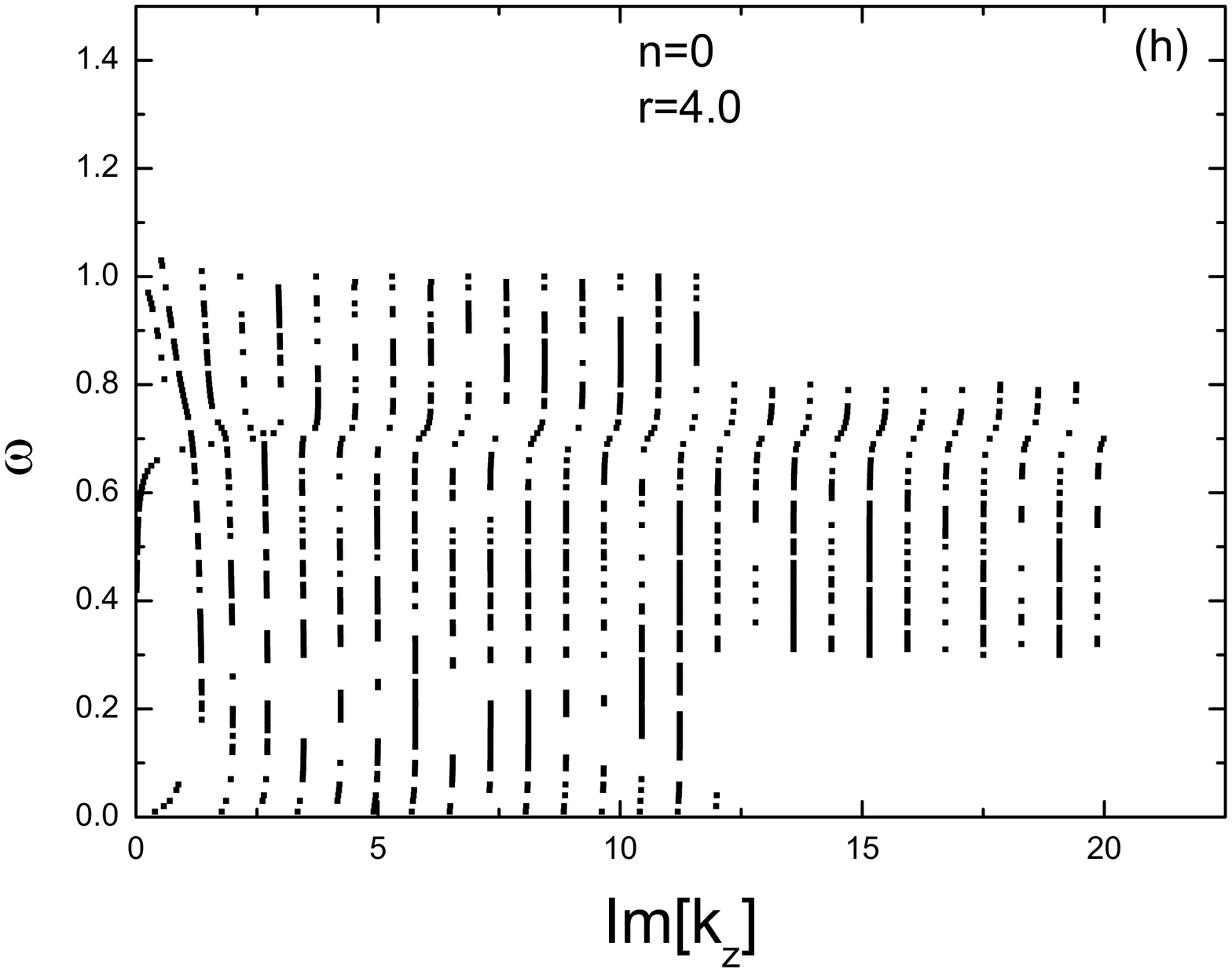}
\caption{$complex-k$ dispersion relations calculated by our method
for cylindrical metallic nanowires with various radius $r$. In the
calculation, $\epsilon_0=1$, $\epsilon_{\infty}=1$, $n=0$ and
$\tau=0.039062$ are used.} \label{fig.4}
\end{figure*}

Before the derivation, three main characteristics of the calculated
BMDs should be clarified. The first characteristic is that the
dispersion curves of the BMDs are all nearly parallel to the
$\omega$ axis, meaning that $Im[k_z]$ of the BMDs are independent to
$\omega$. The second is that the dispersion curves of the BMDs have
a period of $\triangle \ Im[k_z]\times r\approx \pi$. Here,
$\triangle \ Im[k_z]$ is the interval between the curves. The third
characteristic is that the lower dispersion curves with
$\omega<\omega_{sp}$ shift their phases by $\triangle \
Im[k_z]\times r\approx \pi/2$ with respect to the upper curves of
$\omega>\omega_{sp}$.
\begin{figure*}[t!]
\centering
\includegraphics[width=2.0in, height=2.7in, angle=270]{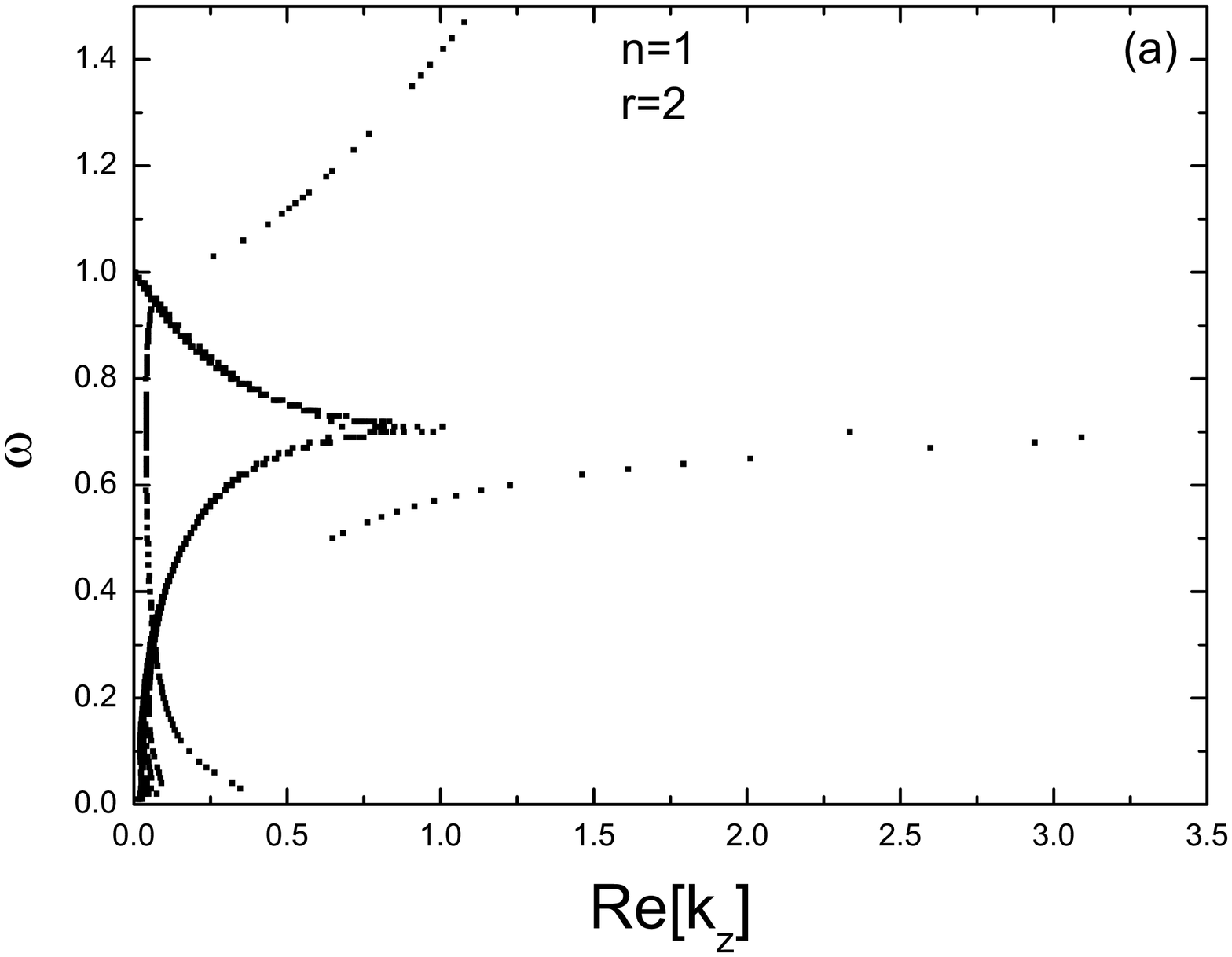}
\hspace*{1cm}
\includegraphics[width=2.0in, height=2.7in, angle=270]{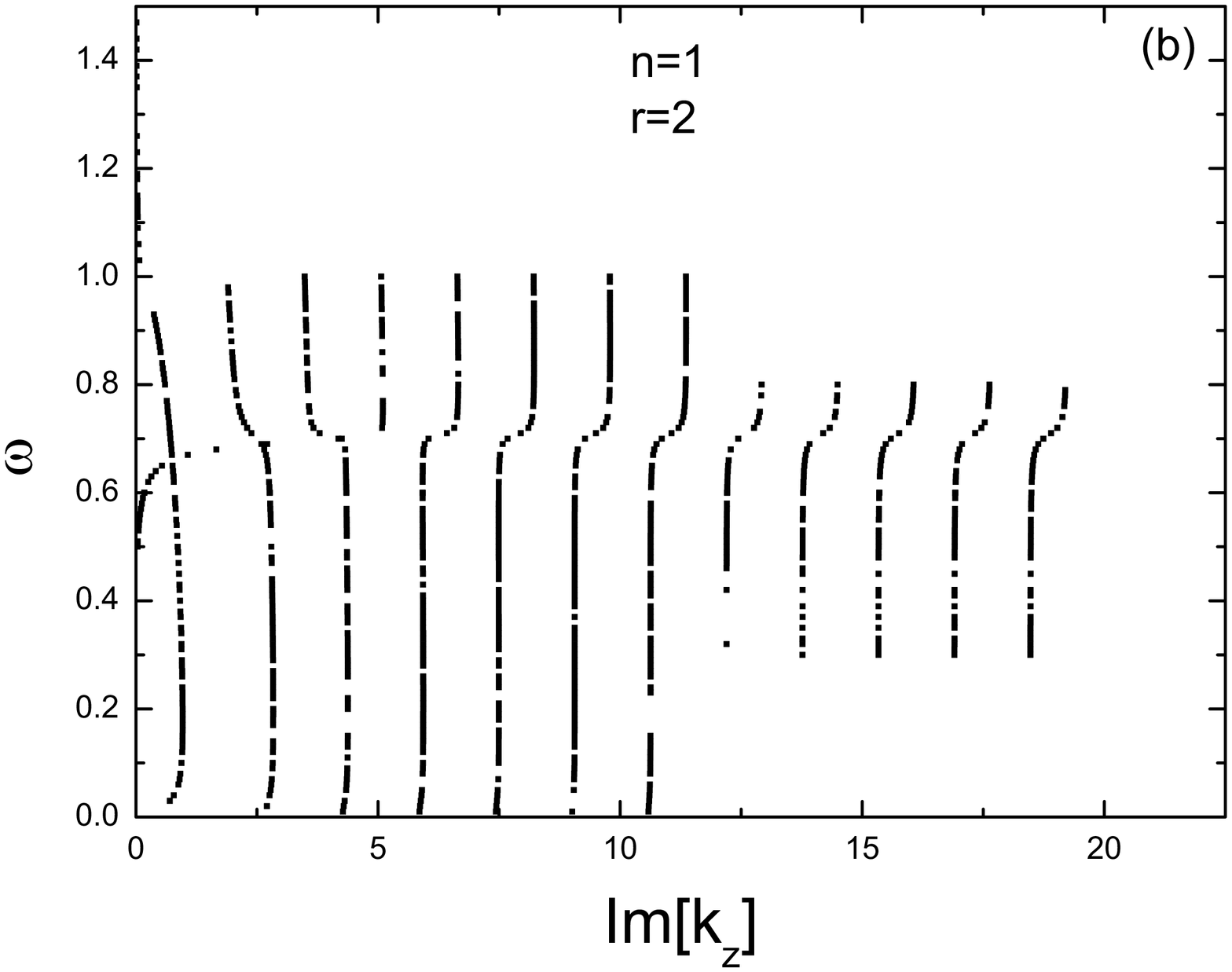}
\\
\includegraphics[width=2.0in, height=2.7in, angle=270]{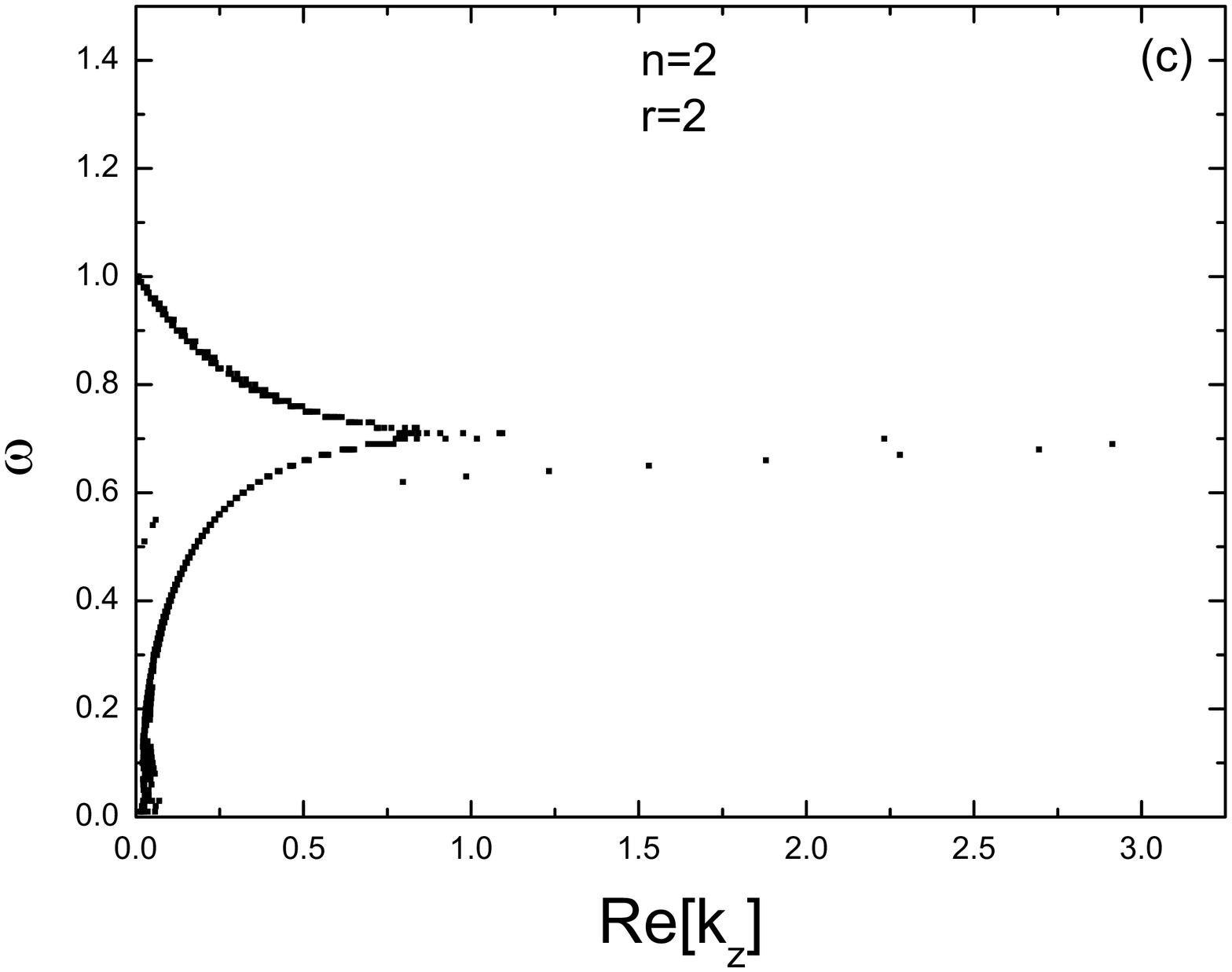}
\hspace*{1cm}
\includegraphics[width=2.0in, height=2.7in, angle=270]{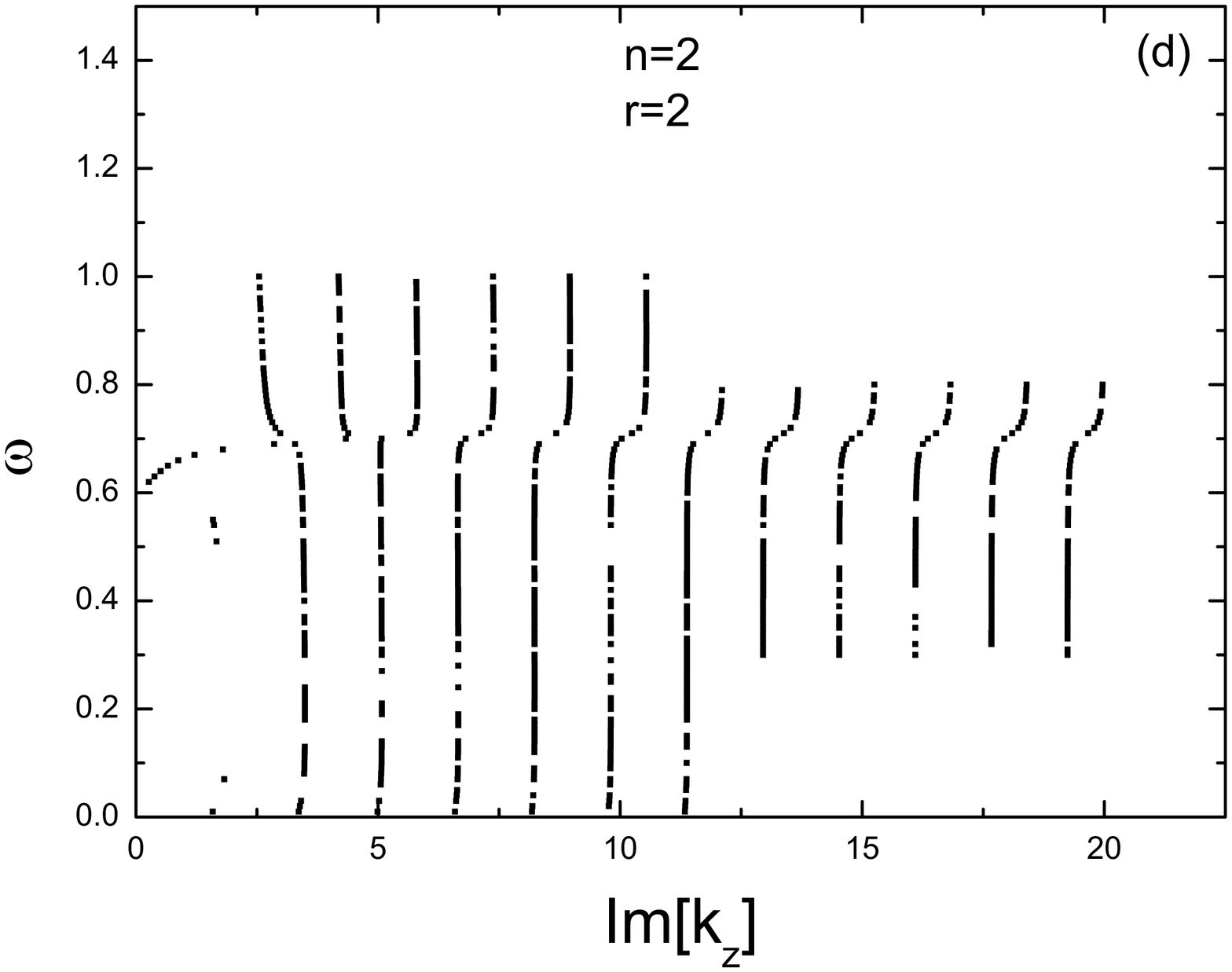}
\\
\includegraphics[width=2.0in, height=2.7in, angle=270]{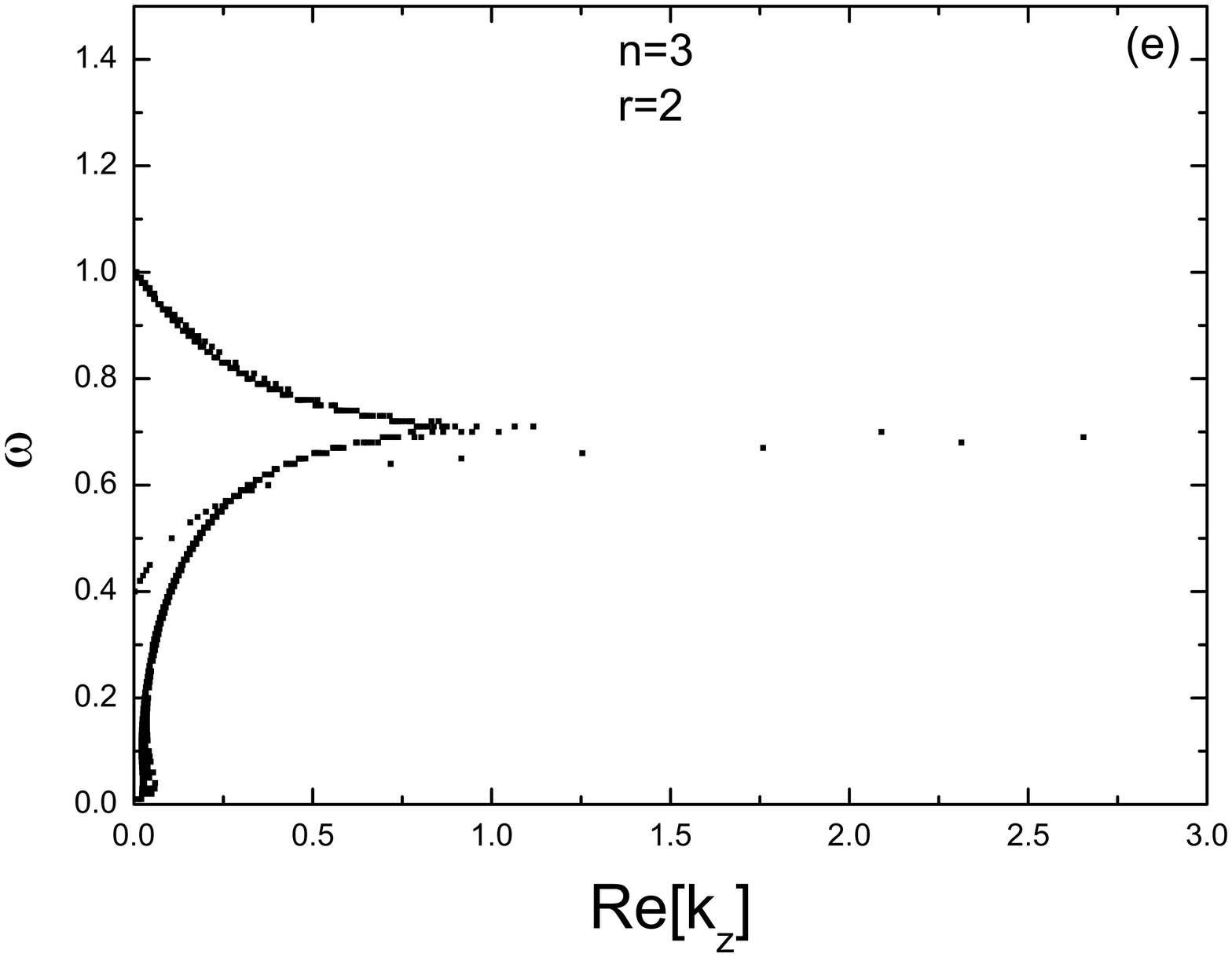}
\hspace*{1cm}
\includegraphics[width=2.0in, height=2.7in, angle=270]{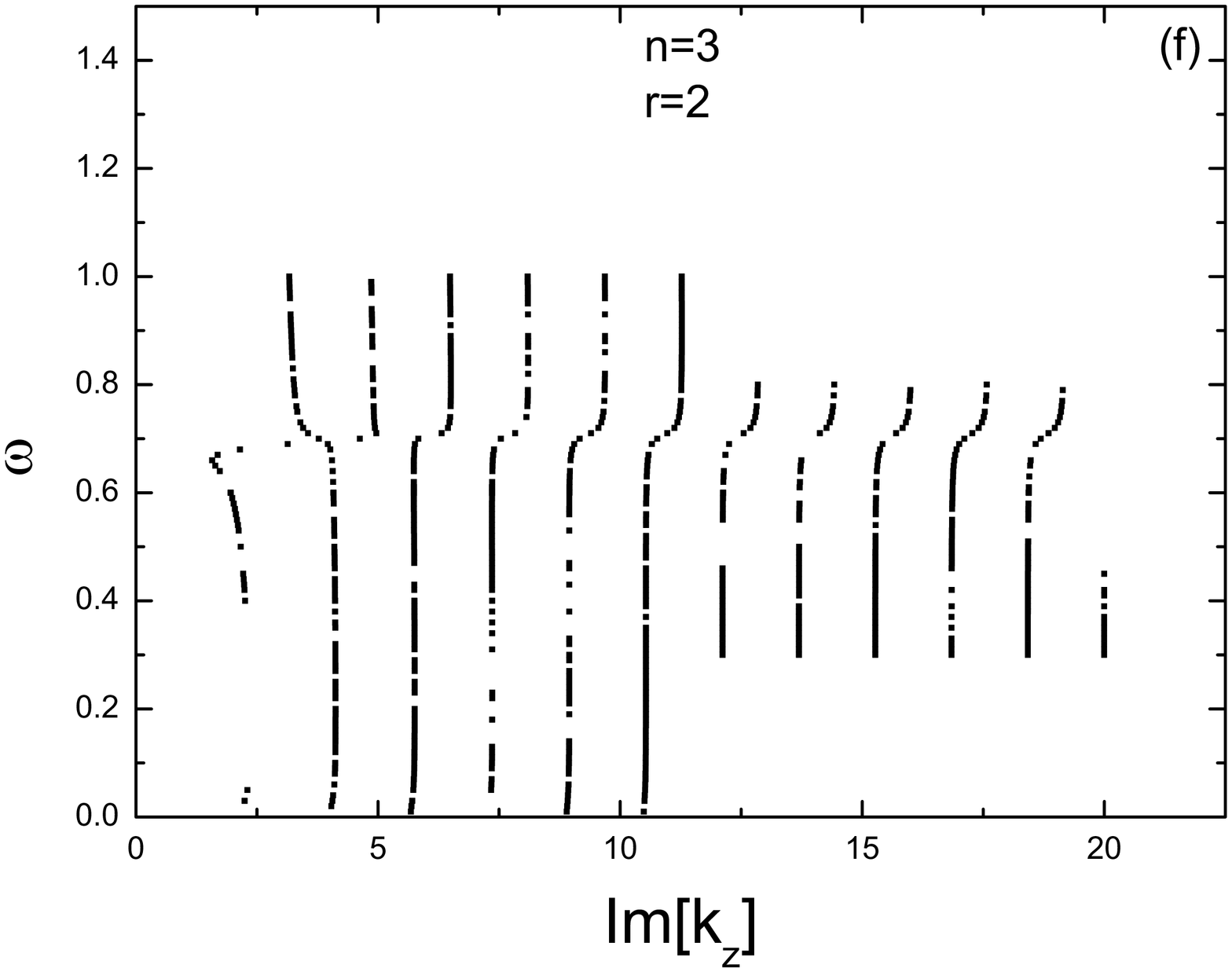}
\\
\includegraphics[width=2.0in, height=2.7in, angle=270]{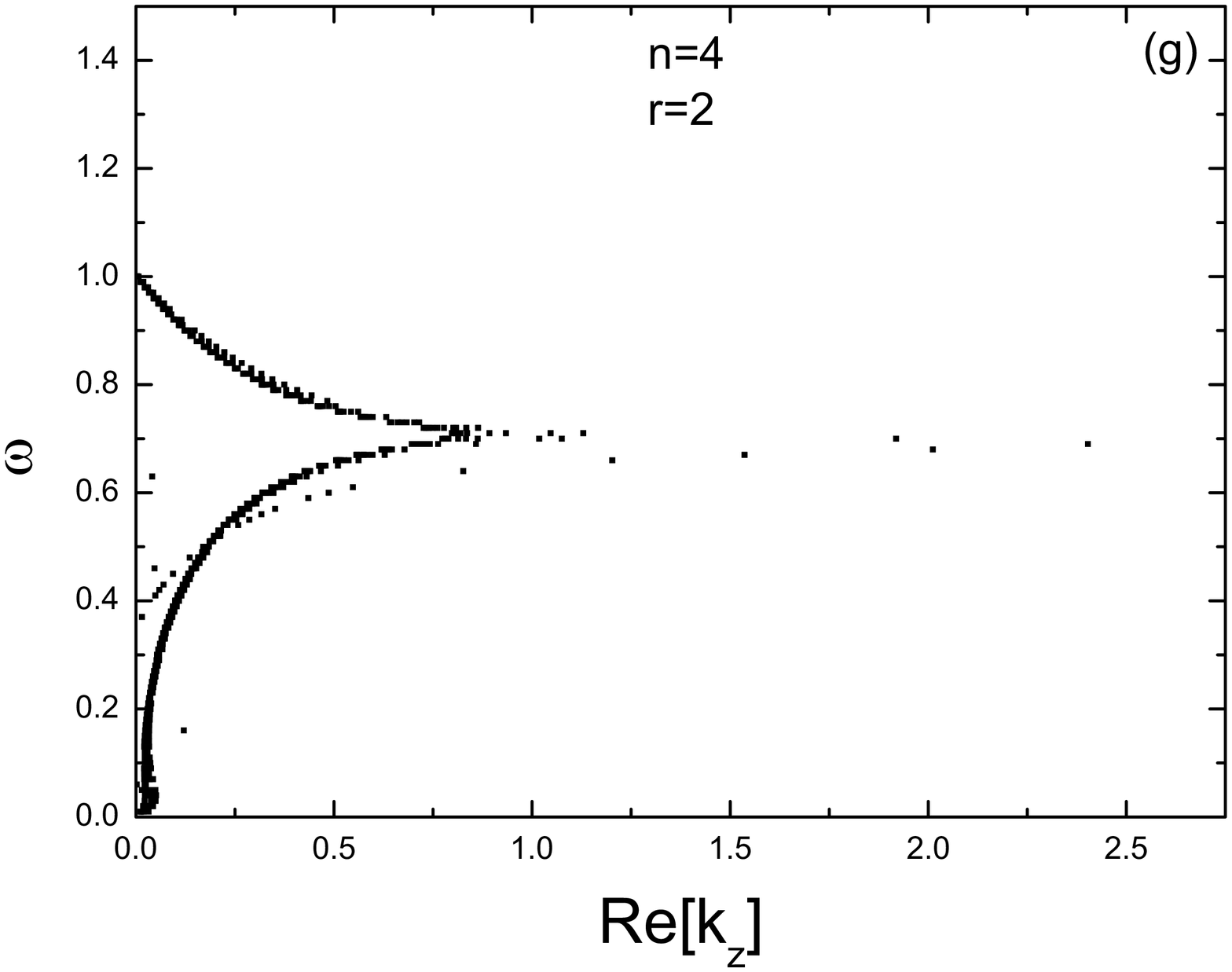}
\hspace*{1cm}
\includegraphics[width=2.0in, height=2.7in, angle=270]{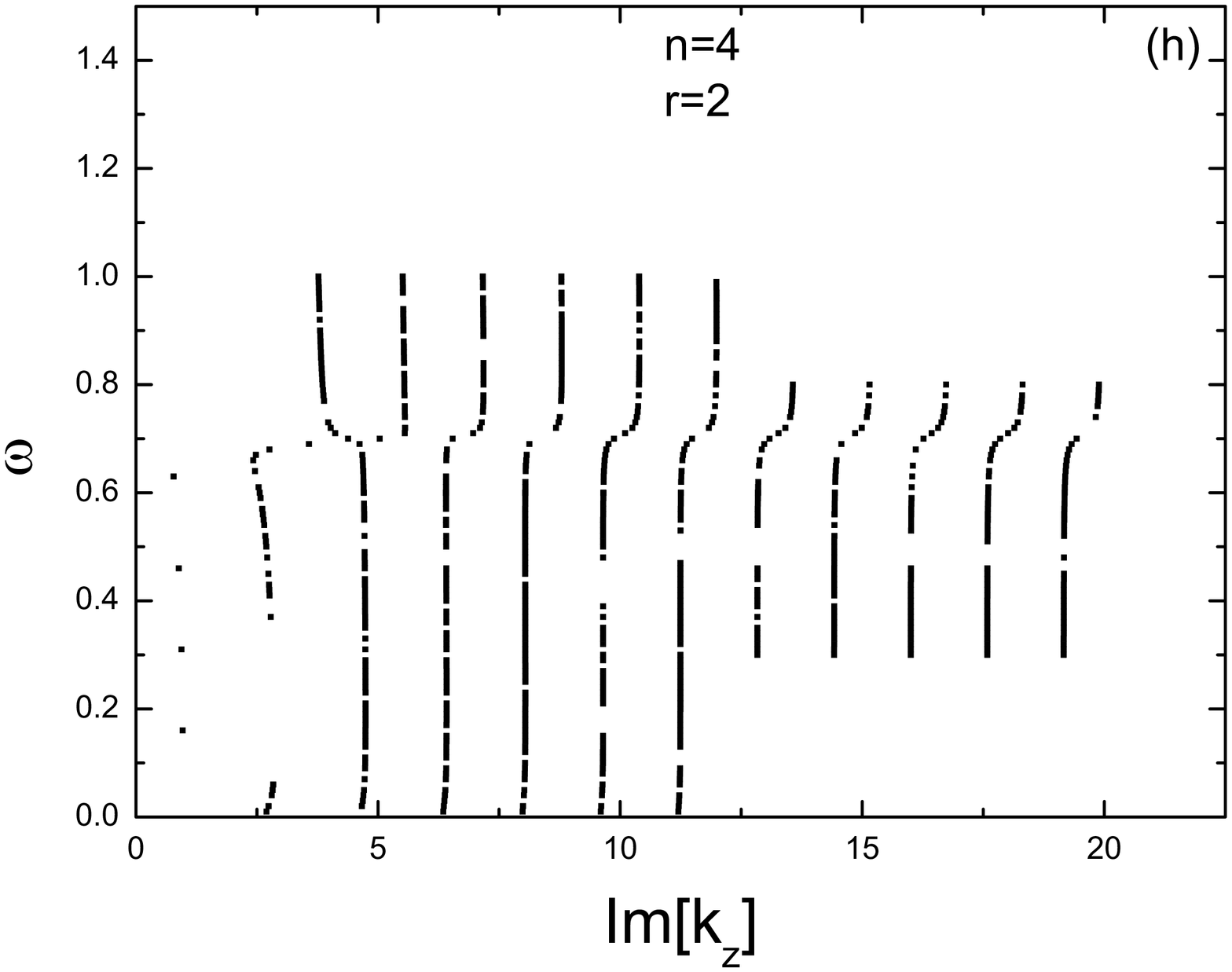}
\caption{$complex-k$ dispersion relations calculated by our method
for a cylindrical metallic nanowire with various orders $n$. In the
calculation, $\epsilon_0=1$, $\epsilon_{\infty}=1$, $r=2$ and
$\tau=0.039062$ are used.} \label{fig.5}
\end{figure*}

For the derivation, it is reasonable for us to consider the limit
case of $Re[k_z]\rightarrow0$ and $Im[k_z]\rightarrow\infty$ based
on fig.~\ref{fig.3}(g) and (h). Thus, the Bessel and Hankel
functions in eq.(\ref{eq.9}) can be replaced by
\begin{subequations}
\label{eq.11}
\begin{equation}
\label{eq.11a} J_0(t)\approx\sqrt{\frac{2}{\pi
t}}cos(t-\frac{\pi}{4}),
\end{equation}
\begin{equation}
\label{eq.11b} H_0^{(1)}(t)\approx\sqrt{\frac{2}{\pi
t}}e^{i(t-\pi/4)}.
\end{equation}
And to get the analytical solution, $k_{rj(j=0,1)}$ should be
expanded by $k_z$ as
\begin{equation}
\label{eq.11c}
k_{rj}\approx\frac{2k_z^2-k_j^2}{2ik_z}.
\end{equation}
 \end{subequations}
For simplicity, the metal is assumed to be perfect with $\tau=0$.
Substituting eqs.(\ref{eq.11})into eq.(\ref{eq.9}), we can get a
simple equation
\begin{equation}
\label{eq.12} e^{-Re[k_z]\times r-i(Im[k_z]\times
r-\frac{\pi}{4})}\approx(\pm)\gamma,
\end{equation}
with
\begin{equation}
\gamma=\frac{1-\omega^2\pm \omega}{\sqrt{1-2\omega^2}}.\nonumber
\end{equation}
For the lower dispersion curves, considering $+\gamma$ at the right
side of eq.(\ref{eq.12}), the complex $k_z$ can be solved
analytically as
\begin{eqnarray}
\label{eq.13}
\left \{
\begin{array}{l}
Re[k_z]=-\frac{\ln (|\gamma|)}{r},\\
\\
Im[k_z]=\frac{\pi/4+2m\pi}{r}.
\end{array}
\right.
\end{eqnarray}
In eq.(\ref{eq.13}), $m$ is an integer. Due to the approximation in
the derivation, the analytical solutions are not coincident to the
calculated results of the BMDs. However, we find that in the
analytical solutions the $Im[k_z]$ is independent to the $\omega$,
which is the first characteristic of the BMDs we calculated.
Secondly, the analytical solution of $Im[k_z]$ shows that the
solutions have a period of $\triangle \ Im[k_z]\times r=2\pi$.
Considering the minus before $\gamma$ in the right side of
eq.(\ref{eq.12}), the period is then equal to $\triangle \
Im[k_z]\times r=\pi$ , which is the second characteristic of the
calculated BMDs results. Last, for the upper curves with
$\omega>1/\sqrt{2}$, the denominator of $\gamma$ is a pure imaginary
value, which is equivalent to the phase shift of $\triangle \
Im[k_z]\times r$  by $\pi/2$ with respect to the lower curves. This
conclusion is just the third characteristic of the BMDs results
calculated by our method. Thus, the approximate analytical solution
to the eq.(\ref{eq.9}) confirms the validity of our method. As shown
in fig.3, full solutions to the complex dispersion equation can be
obtained in once calculation by our proposed method.

In the derivation, $\tau=0$ means that the BMDs are induced by the
structure formation instead of metal loss. For details,
fig.~\ref{fig.4} shows the dispersion curves of BMDs with various
$r$ in the case of $n=0$, $\epsilon_0=1$, $\epsilon_{\infty}=1$, and
$\tau=0.039062$. The curves have two peaks with peak one close to
the surface plamson frequency and peak two at low frequency. For the
nanowires with a smaller radius $r$, the $\triangle \ Im[k_z]$ is
larger to hold the period $\triangle \ Im[k_z]\times r=\pi$. In this
case, peak one is larger while peak two smaller. We also find that
the period is the basic property of the BMDs, which is independent
to the integer order $n$. The BMDs for different orders has been
presented in fig.~\ref{fig.5}, showing the $Im[k_z]$ of BMDs have
the same period but different position.We suggest that the order $n$
only influences the initial position of $Im[k_z]$.

\section{summary}
We have proposed one new method to find full complex roots of a
complex transcendental equation. The correct meshes enclosing the
roots are independent to each other, which guarantee the finding of
the all roots. For the application of this method, the complex
dispersion equation of a cylindrical metallic nanowire is
investigated. In our calculation, locus of the Brewster angle, SPPs
dispersion curves, and bulk modes all can be obtained in once
calculation. Approximate analytical solution to the dispersion
equation has been derived to verify our results. This method can be
applied to all other complex transcendental equations with two real
variables.


\vspace{0.0cm}
\begin{thebibliography}{99}
\vspace{0.5cm}

\bibitem{Barnes1} W. L. Barnes, A. Dereux, and T. W. Ebbesen, Nature(London) {\bf 424}, 824 (2003).

\bibitem{Maier1} S. Maier and H. Atwater, J. Appl. Phys {\bf 98}, 011101 (2005).

\bibitem{Ozbay1} E. Ozbay, Science {\bf 311}, 189 (2006).

\bibitem{Lal1} S. Lal, S. Link, and N. J. Halas, Nature Photon. {\bf 1}, 641 (2007).

\bibitem{Gramotnev1} D. K. Gramotnev and S. I. Bozhevolnyi, Nature Photon. {\bf 4}, 83 (2010).

\bibitem{Talley1} C. E. Talley, J. B. Jackson, C. Oubre, N. K. Grady, C. W. Hollars, S. M. Lane, T. R. Huser, P. Nordlander, and N. J. Halas, Nano Lett. {\bf 5}, 1569 (2005).

\bibitem{Prodan1} E. Prodan, C. Radloff, N. J. Halas, and P. Nordlander, Science {\bf 302}, 419 (2003).

\bibitem{Kelly1} K. L. Kelly, E. Coronado, L. L. Zhao,a nd G. C. Schatz, J. Phys. Chem. B {\bf 107}, 668 (2003).

\bibitem{Abajo1}  F. J. Garcia de Abajo and M. Kociak, Phys. Rev. Lett. {\bf 100}, 106804 (2008).

\bibitem{Chicanne1} C. Chicanne, T. David, R. Quidant, J. C. Weeber, Y. Lacroute, E. Bourillot, A. Dereux, G. Colas des Francs, and C. Girard, Phys. Rev. Lett. {\bf 88}, 097402 (2002).

\bibitem{Ruppin1} R. Ruppin, {\it Electromagnetic Surface Modes} (Wiley, Chichester, 1982).

\bibitem{Archambault1} A. Archambault, T. V. Teperik, F. Marquier, and J. J. Greffer, Phys. Rev. B {\bf 79}, 195414 (2009).

\bibitem{Halevi1} P. Halevi, {\it Electromagnetic Surface Modes} (Wiley, Chichester, 1982).

\bibitem{Rice1}  S. A. Rice, D. Guidotti, and H. L. Lemberg, {\it Aspects of the Study of Surfaces} (Wiley, New York, 1974).

\bibitem{Pfeiffer1} C. A. Pfeiffer, E. N. Economou, and K. L. Ngai, Phys. Rev. B {\bf 10}, 3038 (1974).

\bibitem{Ashley1} J. C. Ashley and L. C. Emerson, Surf. Science {bf\ 41}, 615 (1974).

\bibitem{Chang1} D. E. Chang, A. S. S$\phi$rensen, P. R. Hemmer, M. D. Lukin, Phys. Rev. Lett. {\bf 97}, 053002 (2006);
                 D. E. Chang, A. S. S$\phi$rensen, P. R. Hemmer, M. D. Lukin, Phys. Rev. B {\bf 76}, 035420 (2007).

\bibitem{Chen1} Y. N. Chen, G. Y. Chen, D. S. Chuu, and T. Brandes, Phys. Rev. A {\bf 79}, 033815 (2009).

\bibitem{Novotnv1} L. Novotnv and C. Hafner, Phys. Rev. E {\bf 50}, 4094 (1994).

\bibitem{Arakawa1} E. T. Arakawa, M. W. Williams, R. N. Hamm, and R. H. Ritchie, Phys. Rev. Lett. {\bf 31}, 1127 (1973).

\bibitem{Alexander1} R. W. Alexander, G. S. Kovener, and R. J. Bell, Phys. Rev. Lett. {\bf 32}, 154 (1974).

\bibitem{Udagedara1} Indika B. Udagedara, Ivan D. Rukhlenko, and Malin Premaratne, Phys. Rev. B {\bf 83}, 115451 (2011).

\bibitem{Yao1} Peijun Yao, C. Van Vlack, A. Reza, M. Patterson, M. M. Dignam, and S. Hughes, Phys. Rev. B {\bf 80}, 195106 (2009).

\bibitem{Wan1} Li Wan, Yun-Mi Huang, Chang-Kun Dong, Hai-Jun Luo, arXiv:1108.4797v1.

\end{thebibliography}
\end{document}